%% file: arxivmain.tex
\documentclass[10pt,aps,prb,twocolumn,amsmath,amssymb,floatfix]{revtex4-2}
\input{preamble}

\renewcommand*{\d}{\, \mathrm{d}}      %

\begin{document}

\title{Plasmons in Holographic Ersatz Fermi Liquids}

\author{Eli Ismailov}
\email{ismailov@chalmers.se}
\author{Ulf Gran}
\email{ulf.gran@chalmers.se}
\author{Eric Nilsson}
\email{nieric@chalmers.se}
\affiliation{%
 Department of Physics and Astronomy, Chalmers University of Technology, 41296 Gothenburg, Sweden
}%

\begin{abstract}
We solve an infrared effective holographic model of a non-Fermi liquid at finite temperature that satisfies Luttinger’s theorem and incorporates long-range Coulomb interactions. Motivated by the absence of a Luttinger-counting Fermi surface in standard Reissner-Nordström holographic metals, we consider a Maxwell-Chern-Simons theory in a static anti-de Sitter-Schwarzschild background, coupled to an LU(1) gauge field rather than a conventional U(1) gauge field. By an appropriate choice of boundary conditions, we obtain a damped collective plasmon mode whose plasma frequency scales as predicted by Luttinger's theorem. We further analyze the density-density correlator in the absence of long-range Coulomb interactions and identify a contribution consistent with a Lindhard-like continuum.
\end{abstract}

\maketitle

\section{\label{introduction}Introduction}

Strongly correlated metallic states that lack long-lived quasiparticles remain a central challenge in condensed matter physics. Prominent examples include strange metals, whose anomalous transport and spectroscopic properties often lie outside the scope of conventional Fermi-liquid theory. 
Holographic duality has emerged as a useful framework for studying such systems, and which maps certain strongly interacting quantum many-body theories to classical gravitational dynamics in one higher dimension~\cite{zaanen_holographic_2015,hartnoll_holographic_2018}. 
Such approaches have provided tractable models exhibiting, for example, linear-in-temperature resistivity~\cite{balm_t-linear_2023}
as well as momentum-scaling nodal ARPES spectral functions \cite{smit_momentum-dependent_2024}, 
and provide complementary perspectives on solvable strongly interacting models such as SYK systems.

In the context of compressible metallic phases, however, conventional holographic constructions face an important conceptual difficulty. 
Finite charge density is typically introduced through charged black hole geometries. Most commonly, these are Reissner-Nordström backgrounds, which have a bulk electric charge residing behind the black hole horizon. This implies a Fermi surface in the boundary theory that does not account for the full microscopic charge density~\cite{iqbalLuttingersTheoremSuperfluid2012}, leading to an apparent violation of Luttinger's theorem~\cite{luttinger_fermi_1960} due to missing charge \footnote{There are holographic electron star models where the charge density is sourced explicitly by bulk fermions, c.f.~Refs.~39-41. However, these models tend to have a large number of smeared Fermi surfaces, making them less relevant for typical condensed matter scenarios.}. 
Although quantum corrections can, in principle, restore the correct counting \cite{Polchinski_2012, faulknerFriedelHorizon2013}, this undermines the original appeal of holography as a classical dual description. These issues indicate that conventional holographic models do not yet provide a satisfactory description of metallic phases with a well-defined Fermi surface.

Recent work by Else, Thorngren and Senthil (ETS) \cite{else_non-fermi_2021} introduced a different perspective on metallic phases through the framework of \textit{ersatz Fermi liquids} (EFLs) \cite{else_strange_2021, elseCollisionless2023, else_critical_2021, else_t_2025, lu_definition_2024, huang2024Effectivea}, which incorporates both Fermi liquids and non-Fermi liquids. In this framework, metallic phases are characterized by emergent infinite-dimensional symmetry groups with specific 't Hooft anomalies. The Fermi surface is encoded through the anomaly structure, rather than dynamically through long-lived quasiparticle excitations. 

The emergent symmetry strongly constrains the infrared theory and leads to, inter alia, a kinematic formulation of Luttinger’s theorem, making it insensitive to the strength of the interactions.
The EFL framework therefore describes the IR theory of both Fermi and non-Fermi liquids, as long as they are metals with closed Fermi surfaces. Any consistent model of a strongly correlated clean compressible (non-)Fermi liquid should therefore satisfy the structural constraints implied by EFLs, including Luttinger’s theorem.

Ref.~\cite{else_holographic_2024} proposed a holographic realization of an EFL that indeed satisfies Luttinger's theorem, and which could be used as a platform for further studies extending the holographic framework towards more realistic models of strange metals.
In that construction, the bulk geometry may remain charge neutral, while the boundary charge density is encoded through the emergent anomalous symmetry structure rather than by electric flux through a black hole horizon.

In this paper, we extend this construction by incorporating long-range Coulomb interactions, which qualitatively modify the charge response and lead to plasmon excitations. This is achieved by gauging the anomaly-free subgroup of the emergent symmetry through a choice of mixed boundary conditions, resulting in a dynamical boundary gauge field.
We then compute the finite-temperature density response of the dual field theory, where our main result is that the spectral weight indeed reorganizes into a damped collective plasmon mode with a finite gap at zero momentum.
The plasma frequency scales as $\omega_p\propto \sqrt{k_{\mathrm F}} $, which we show is in agreement with Luttinger’s theorem. Using instead the standard setup of a non-dynamical, external gauge field yields a continuum reminiscent of a Lindhard particle-hole spectrum.
Together, these results show that collective charge dynamics characteristic of metals can emerge in a holographic theory whose charge sector is encoded through the anomaly structure rather than conventional finite-density bulk flux. The damping of the plasmon remains at zero momentum, reflecting that the quantum critical continuum characteristic of holographic theories remains.

The remainder of the paper is structured as follows. In~\cref{sec:ErsatzFermiLiquids} we briefly outline the ersatz Fermi liquid framework, the emergent LU(1) symmetry and its associated ’t Hooft anomaly kinematically encoding the Fermi surface~\cite{else_non-fermi_2021}.
In \cref{sec:HolographicDual} we extend the holographic construction of \cite{else_holographic_2024} to finite temperature deriving the linearized bulk equations governing charge fluctuations. In \cref{sec:BoundaryConditions} we discuss the boundary conditions implementing dynamical electromagnetism and in \cref{sec:Results} we present the numerically computed density-density response describing a Lindhard-like continuum and the plasmon. We conclude in \cref{sec:ConclusionAndOutlook} with a discussion of implications and future directions. Technical details are presented in three appendices.

\section{Ersatz Fermi liquids}\label{sec:ErsatzFermiLiquids}
In this section we briefly review the framework of ersatz Fermi liquids developed in Ref.~\cite{else_non-fermi_2021}, which provides a symmetry-based characterization of metallic phases.
Within this formulation, a metallic phase is taken as being compressible, meaning that the microscopic filling can vary continuously. In addition, the system is assumed to be clean, possessing continuous or lattice translational symmetry $\mathbb{R}^d$ or $\mathbb{Z}^d$.
A central distinction is made between kinematic and dynamical properties. Kinematic properties are universal and determined by structural features of the theory such as symmetries, anomalies, and topological constraints. They characterize the structure of the Hilbert space. Dynamical properties, on the other hand, depend on the specific interactions and couplings defining the theory and determine the dynamical evolution of states within this Hilbert space.
Although Fermi liquids (FL) and many non-Fermi liquids (NFL) differ greatly in their dynamics, they share the same infrared kinematics. Consequently, any result derived purely from these kinematic properties must hold for all members of the same EFL class.

A defining kinematic feature of an EFL is the emergence of an infinite-dimensional internal symmetry group in the infrared, denoted $G_\text{IR}$, emergent from a microscopic symmetry under renormalization group (RG) flow. The symmetry group $G_\mathrm{IR}$ for metallic EFLs is given as the loop-$F$ group of $G$, denoted L$_F G$. This symmetry expresses the conservation of $G$-charge density locally along a closed manifold $F$, which up to a homeomorphism can be identified with the Fermi surface. Alternatively, we can take $F$ to provide an abstract definition of a Fermi surface that remains meaningful even for non-Fermi liquids.

From this point onward we specialize to the case of a two-dimensional metal with conserved U(1) charge. We may therefore take \footnote{Any one-dimensional closed manifold is homeomorphic to a disjoint union of circles. We restrict to the most parsimonious case of $F = S^1$.} $F = S^1$, which we parametrize with an angle $\theta$. The emergent symmetry is thus ${G_{\mathrm{IR}} = \mathrm{L}_{S^1}\mathrm{U}(1) \equiv \mathrm{LU}(1)}$, the group of smooth maps ${S^1 \to \mathrm{U}(1)}$.

Physically, this symmetry corresponds to the existence of infinitely many conserved charges generated by the operator-valued distribution $n(\theta)$, describing the $\theta$-resolved charge density along the Fermi surface. Associated with this conserved charge density we can define a conserved current through the continuity equation
\begin{equation}
    n(\theta) = \int \d^2 x \; j^0(x^\alpha,\theta)\,, \qquad \partial_\alpha j^\alpha = 0\,,
    \label{eq:lu continuity eq}
\end{equation}
where $\alpha=t,x,y$ runs over the space-time coordinates only, i.e.,~not including $\theta$.
In a conventional Fermi liquid this structure arises because scattering processes that transfer quasiparticles between distinct points on the Fermi surface are irrelevant in the renormalization-group sense, leaving only forward-scattering Landau interactions as marginal couplings in the infrared~\cite{shankar_renormalization-group_1994,polchinski_effective_1999}.
Since the EFL framework takes the resulting symmetry structure itself to be fundamental, it should persist even when quasiparticles are absent.

To probe the LU(1) symmetry, one couples the theory to a background LU(1) gauge field. An infinitesimal local symmetry transformation is parametrized by a function $\lambda(x,\theta)$, which results in a gauge field $A_\alpha(x,\theta)$ transforming as
\begin{equation}
    A_\alpha(x^\alpha,\theta) \mapsto A_\alpha(x^\alpha,\theta) + \partial_\alpha \lambda(x^\alpha,\theta)\,.
\end{equation}
Because the symmetry is defined along the Fermi surface, the gauge field also contains a $\theta$-component,
\begin{equation}
    A_\theta(x^\alpha,\theta) \mapsto A_\theta(x^\alpha,\theta) + \partial_\theta \lambda(x^\alpha,\theta)\,.
\end{equation}
An LU(1) gauge field on a (2+1)-dimensional space-time $M_3$ can therefore be viewed as a U(1) gauge field on the product manifold $M_3 \times S^1$ with components $A_A(x^\alpha,\theta)$, where the index $A=(\alpha,\theta)$ labels both the space-time directions $\alpha$ of $M_3$ and the internal $S^1$ direction. The corresponding gauge transformation
\begin{equation}
    A_A(x^\alpha,\theta) \mapsto A_A(x^\alpha,\theta) + \partial_A \lambda(x^\alpha,\theta)
\end{equation}
thus takes the form of a U(1) gauge field on $M_3 \times S^1$.
In a conventional Fermi liquid, the additional component $A_\theta$ may be interpreted as encoding the Berry phase accumulated by a quasiparticle transported along the Fermi surface \cite{else_holographic_2024}. 

In the presence of background gauge fields, the conservation of charge density at each $\theta$ breaks down. The LU(1) current of \cref{eq:lu continuity eq} is therefore no longer conserved, but instead obeys the 't Hooft anomaly equation
\begin{equation}
    \partial_A j^A = \frac{m}{8\pi^2} \epsilon^{ABCD}\partial_A A_B \partial_C A_D\,,
    \label{eq:tHooft_anomaly_general}
\end{equation}
where we have allowed for a $j^\theta$-component of the current and $m \in \mathbb{Z}$ is an anomaly coefficient. Physically, this anomaly reflects the fact that charge localized at a point on the Fermi surface can be transported along the surface when the system is subjected to electromagnetic fields. As will be relevant for the holographic formulation discussed in \cref{sec:HolographicDual}, the anomaly can alternatively be understood via anomaly inflow from a (3+1)-dimensional manifold $M_4$, with boundary $\partial M_4 = M_3$, from the action
\begin{equation}
    S_{\mathrm{CS}}=\frac{m}{24\pi^2}\int_{M_4\times S^1} A\wedge \d A\wedge \d A \,,
    \label{eq:anomaly_inflow}
\end{equation}
where the $S^1$ is the internal Fermi surface direction. The gauge field variation evaluated at the boundary then gives a current,
\begin{equation}
    j^A_\text{CS}=\frac{\delta S_{\mathrm{CS}}}{\delta A_A}\Big |_{\partial M_4}= \frac{m}{8\pi^2} \epsilon^{ABCD}A_B \partial_C A_D~,
    \label{eq:inflow_current}
\end{equation}
reproducing the right hand side of the 
anomaly
equation \cref{eq:tHooft_anomaly_general}.

\subsection{Luttinger's theorem}\label{sec:LuttingersTheorem}
As shown by ETS~\cite{else_non-fermi_2021}, Luttinger’s theorem can be formulated in a purely kinematic way. 
A central result of Ref.~\cite{else_non-fermi_2021} is that Luttinger's theorem can be formulated purely in terms of infrared symmetry data. Once three ingredients are known; 
\begin{enumerate}
    \item the emergent symmetry $G_\text{IR}$,
    \item the homomorphism ${\varphi:\mathbb{R}^d \times \text{U}(1) \rightarrow  G_\text{IR}}$ that specifies how microscopic translations and charge symmetry are realized in the IR,
    \item the 't Hooft anomaly of the IR theory,
\end{enumerate}
the microscopic charge density $\rho$ is fixed, independent of the interaction strength.

For the present case of a two-dimensional LU(1) metal, Luttinger's theorem reads
\begin{equation}
    \rho \;=\; \frac{m\mathcal{V}_\text{F}}{(2\pi)^2},
    \label{eq:generalised-LT}
\end{equation}
where $\mathcal{V}_\text{F}$ is the volume enclosed by the Fermi surface. The anomaly coefficient, $m\in\mathbb{Z}$, corresponding to the Chern-Simons level of \cref{eq:anomaly_inflow}, captures the multiplicity of the Fermi surface \cite{lu_definition_2024}.

Luttinger's theorem thus follows directly from the anomaly structure of the emergent LU(1) symmetry, making the Fermi volume tied to the microscopic charge density independently of coupling or dynamics. The significance for the present work is immediate:
any holographic model realizing the LU(1) symmetry with the correct anomaly structure should therefore automatically satisfy Luttinger’s theorem.

\section{Holographic dual}\label{sec:HolographicDual}
We now move on to the holographic effective theory used to model the ersatz Fermi liquid introduced in the previous section. This construction should be viewed as an IR effective holographic model, capturing the universal low-energy structure associated with the emergent symmetry and its anomaly, rather than a UV-complete description of the microscopic theory. Our starting point is the Maxwell-Chern-Simons construction proposed in Ref.~\cite{else_holographic_2024}, which realizes the emergent LU(1) symmetry and its associated anomaly in a classical bulk description. Here, we consider that framework at finite temperature and formulate the linear-response problem relevant for collective charge dynamics.

Introducing a higher dimension and evaluating the boundary currents through the anomaly-inflow formulation \cref{eq:inflow_current} is reminiscent of the holographic treatment of boundary theories. Holography is therefore used to model a strongly correlated effective theory with a global LU(1) symmetry and the correct anomaly by introducing a dynamic anomaly-free LU(1) gauge field in a bulk spacetime equipped with a $5$-dimensional Chern-Simons action term. The global symmetry current on the boundary of $M_4\times S^1$ then holographically corresponds to a theory with the 't Hooft anomaly for the LU(1) symmetry as described above, while, as is standard to any holographic treatment, the duality ensures strong correlations. 

In addition to the Chern-Simons term, a bulk theory containing an LU(1) gauge field requires a Maxwell-term. The bulk action dual to the EFL thus becomes 
\begin{equation}
    \begin{aligned}
        S_\mathrm{M} &+ S_\mathrm{CS} + S_\mathrm{EH} = \\ 
        & -\frac{1}{4} \int_{M_4} \text{d}^4x \sqrt{-g} \int_{S^1} \text{d}\theta \frac{1}{\alpha(\theta)} F_{\mu\nu}(\theta)F^{\mu\nu}(\theta) \\
        & +\frac{m}{24\pi^2}\int_{M_4\times S^1} \ A\wedge \d A\wedge \d A \\
        & + S_\mathrm{EH}|_{M_4}\,,
        \label{eq:bulk action}
    \end{aligned}
\end{equation}
where $M_4$ denotes the four-dimensional bulk spacetime, $S^1$ is the direction of the internal Fermi surface coordinate $\theta$, and $S_\mathrm{EH}$ is the Einstein-Hilbert term determining the geometry, which is asymptotically anti-de Sitter (AdS) \footnote{AdS is a maximally symmetric solution to Einstein's equations with a negative cosmological constant.}. Note that the latter does not contain the internal $S^1$ direction. The indices $\mu,\nu$ run over the bulk spacetime directions only (excluding $\theta$), $m$ is the Chern-Simons level giving the anomaly coefficient discussed above, and $\alpha(\theta)$ acts as a $\theta$-dependent gauge coupling.

The standard holographic treatment is to consider the gauge field variation of the bulk action at the conformal boundary to identify the dual currents. In the presence of a Chern-Simons term, the gauge field variation gives a so-called consistent current, which can be split into a covariant part and a Bardeen-Zumino term \cite{bardeen1984Consistent}. The transport properties of the system are then given by the covariant current~\cite{gynther2011Holographic, jimenez-alba2014Anomalous, landsteiner2012Holographic}, which, in this case, coincides with the variation of the Maxwell action alone. The observable covariant current of interest to us is therefore given by
\begin{equation}
    j^A \equiv j^A_\text{cov} =\frac{\delta S_{\mathrm{M}}}{\delta A_A}\Big|_{\text{AdS boundary}}~,
    \label{eq:current}
\end{equation}
where the reduction to the boundary contribution is due to being on-shell. Crucially, $A_\theta$ is a fixed background,
 which ensures that $j^\theta=0$ on the boundary, a condition necessary to truly have LU(1)-symmetry, as $j^\theta\neq0$ would imply dynamic non-forward scattering~\cite{else_holographic_2024}.

The metric is thus four-dimensional, with the $\theta$-direction acting as an index for an internal symmetry. The Maxwell term can be interpreted as a family of gauge theories on $M_4$ parametrized by the coordinate $\theta$, or equivalently as a sum over internal ``flavor'' sectors distinguished by $\theta$. 

\subsection{Finite-temperature background}
The equilibrium background solution of \cite{else_holographic_2024} describes a static Fermi surface. To see how it appears, consider the anomalous continuity equation \cref{eq:tHooft_anomaly_general}, but write it in the form
\begin{equation}
    \partial_A j^{A} = \frac{m}{8 \pi^2}\left(2 \epsilon^{ij} E_i F_{\theta j} + \frac{1}{2} B F_{\theta t} \right)\,.
\end{equation}
This may be contrasted with the 't Hooft anomaly from Fermi liquid theory \cite{else_holographic_2024}, which, for ${B=0}$, $E\neq0$, reads
\begin{equation}
    \partial_A j^{A} = \frac{m}{(2\pi)^2} \epsilon^{ij} E_i F_{\theta j} \overset{\text{FLT}}{=} \frac{1}{(2\pi)^2} \epsilon^{ij} E_i \partial_\theta k_j(\theta)\,,
    \label{eq:FLTanomaly}
\end{equation}
where $i,j$ are spatial indices. The Fermi surface appears as $k_i(\theta)$ in momentum space, parametrized by $\theta$. Thus, in a Fermi liquid $F_{\theta i} = \partial_\theta k_i(\theta)$, and with an appropriate gauge choice, $\partial_\theta\left[A_i - k_i(\theta)\right] = 0$ identifies $A_i$ as the Fermi surface up to an additive constant independent of $\theta$. This relation is taken to be the definition of the Fermi surface also for a non-Fermi liquid.

Varying the action \cref{eq:bulk action} results in the Einstein and Maxwell-Chern-Simons equations of motion
\begin{align}
    R_{\mu \nu}-\frac{1}{2} g_{\mu \nu}\left(R + 2 \Lambda \right) &= F_{\mu \kappa} F^\kappa_\nu -\frac{1}{4} g_{\mu \nu} F_{\kappa \lambda} F^{\kappa \lambda}\,, \\
    \partial_\nu[\sqrt{-g}F^{\nu \mu}] &= \frac{m\alpha(\theta)}{(2\pi)^2} \epsilon^{\mu \nu \rho \sigma} \partial_\theta A_\nu \partial_\rho A_\sigma\,,
\end{align}
where $\Lambda = -3/L^2$ in four dimensions. For finite density matter, one normally considers charged black hole (Reissner-Nordström) solution which introduces a microscopic chemical potential $\mu$ at the UV boundary, where the gauge field component $A_t$ captures the full RG flow of the charge density that sources it. The holographic EFL model considered here and in \cite{else_holographic_2024} only attempts to describe the effective theory in the deep IR, which is a Fermi surface at zero charge density. The background solution should therefore only contain the Fermi surface in the absence of a chemical potential sourcing a charge density.

This means that we can consider backgrounds of the form
\begin{equation}
    \begin{aligned}
        &\d s^2 = \frac{L^2}{z^2}\left(-f(z)\d t^2 + \d x^2 + \d y^2 + \frac{1}{f(z)}\d z^2\right)\,,\\
        & \bar A_i = k_{i}(\theta), \quad 
        \bar A_t=
        \bar A_z = 0\,,
    \end{aligned}
    \label{eq:metric}
\end{equation}
i.e., a static AdS-Schwarzschild black hole and two spatial gauge field components encoding the Fermi surface. Because the components of the gauge field only depend on $\theta$, they do not enter the stress tensor on the right-hand side of Einstein's equations, which makes the Schwarzschild solution consistent.

The emblackening factor takes the form
\begin{equation}
    f(z)=1-z^3\,,
\end{equation}
where we set the horizon at $z=1$ without loss of generality.

\subsection{Linearized equations of motion}\label{sec:linearEOM} To obtain the response functions of the boundary theory, the gauge field is perturbed as
\begin{equation}
    A_\mu(x^\mu, \theta) \mapsto \bar{A}_\mu(\theta) + a_\mu(x^\mu, \theta)
\end{equation}
around the static background $\bar{A}_\mu(\theta)$ in \cref{eq:metric}.

Linearizing the bulk equations of motion yields a set of coupled differential equations for the fluctuations $a_\mu(x^\mu,\theta)$, where we 
add a Lorenz-type gauge fixing term to render the radial boundary value problem numerically well posed:
\begin{equation}
    \begin{aligned}
            \partial_\nu[\sqrt{-g}f^{\nu \mu}]  + \partial^\mu \bigg[ \frac{1}{\sqrt{-g}}&\partial_\nu(\sqrt{-g}a^\nu) \bigg] \\
            &= \frac{m\alpha(\theta)}{(2\pi)^2} \epsilon^{\mu \nu \rho \sigma} \partial_\theta \bar A_\nu \partial_\rho a_\sigma\,,
    \end{aligned}
    \label{eq:BulkEOM_lin_gf}
\end{equation}
where $f = \d a$. Verifying that the Lorenz-type gauge term in square brackets is numerically zero serves as a non-trivial check of our numerics.
Note that the linearity of \cref{eq:BulkEOM_lin_gf} implies that any coupling to dynamic gravity at the level of linear response is impossible without modifying the background. Furthermore, due to the background solution \cref{eq:metric}, the Einstein equations receive no contributions from the gauge field fluctuations to linear order, since they are quadratic in $F$. We can thus consistently set the metric fluctuations to zero.  

Since the background is translationally invariant in the boundary spacetime directions, we may take the perturbations in plane-wave form,
\begin{equation}
    a_\mu(x^\mu, \theta)\mapsto a_\mu(z, \theta) \mathrm{e}^{-\mathrm{i} \omega t + \mathrm{i} k x}\,.
\end{equation}
The end result is four linear ordinary differential equations in the radial direction $z$, whose explicit form can be found in~\cref{app:EOMs}.

\section{Boundary conditions}\label{sec:BoundaryConditions}
With the bulk equations of motion in place, what remains are the boundary conditions that are crucial for
incorporating long-range Coulomb interactions.
The first boundary is at the horizon of the AdS-Schwarzschild black hole, where $z=1$. To select modes that carry energy toward the black hole, corresponding to causal response functions in the boundary theory, 
we pull out an infalling factor from the fluctuations $a_\mu(z,\theta)$, as is further detailed in \cref{app:horizonBC}.

At the conformal boundary, $z=0$, the boundary conditions specify which sources are held fixed in the dual quantum field theory and therefore determine the $n$-point functions that are computed. In the present case these are the electromagnetic current-current response functions of an ersatz Fermi liquid describing collective charge dynamics and transport.

\subsection{Conformal boundary conditions}
Whether the total boundary charge is dynamical or not is determined by the choice of boundary conditions at the conformal boundary. Conventionally, one sets Dirichlet boundary conditions for the bulk gauge field, which corresponds to a fixed source for a globally conserved dual current. Alternatively, a specific choice of mixed, Robin-type boundary conditions promotes the boundary photon to be dynamical, gauging the U(1) and thus implementing the long-range Coulomb interactions between charges~\cite{gran_holographic_2018}. 

In the present case, this procedure is not as clear
since the 't Hooft anomaly prevents the gauging of the entire LU(1) group. Instead, only the anomaly-free total-charge subgroup may be promoted to a gauge symmetry. 

The identification of this subgroup is found in the group isomorphism
\begin{equation}
    \mathrm{LU(1)} \cong \; \mathrm{U(1)} \times \Omega^{(0)}_1\mathrm{U(1)} \times \mathbb{Z}\,,
    \label{eq:fullLU1split}
\end{equation}
where U(1) is the set of constant maps, $\mathbb{Z}$ specifies the winding of loops around U(1) and $\Omega^{(0)}_1\mathrm{U(1)}$ the zero winding component of based loops to U(1). The Lie algebra of LU(1) can canonically be identified as $\mathrm{L}\mathbb{R}$ wherefrom we find a Lie algebra isomorphism
\begin{equation}
    \mathrm{L}\mathbb{R} \cong \mathbb{R}\oplus\Omega_0 \mathbb{R}\,,
    \label{eq:Lu1split}
\end{equation}
where $\mathbb{R}$ is the Lie algebra of U(1) and $\Omega_0 \mathbb{R}$ the Lie algebra of $\Omega^{(0)}_1\mathrm{U(1)}$. A more detailed discussion of these relations can be found in~\cref{app:loopgroup}.
With this split of the loop group we offer a physical interpretation: there is an infinite set of LU(1) gauge fields corresponding to an infinite set of generators. With $\mathrm{L}\mathbb{R}$  being real periodic functions, we can formally Fourier expand the gauge fields and associate each mode with the generators of the algebra in \cref{eq:Lu1split}. Specifically, the zero mode corresponds to $\mathbb{R}$ which is the generator of U(1) in \cref{eq:fullLU1split}. As mentioned, it is this U(1) that we wish to gauge.

Following this analysis we consider the boundary currents, decomposed into Fourier modes, 
\begin{equation}
    j^{\alpha}(\theta, x^\alpha) \;=\; j^{\alpha}_{(0)}(x^\alpha) \;+\; \sum_{n \neq 0} j^{\alpha}_{(n)}(x^\alpha)\, e^{\mathrm{i} n \theta}\,,
    \label{eq:fourier_current}
\end{equation}
sourced by their respective Fourier mode of a background gauge field,
\begin{equation}
    A_{\alpha}(\theta, x^\alpha) \;=\; A_{\alpha}^{(0)}(x^\alpha) \;+\; \sum_{n \neq 0} A_{\alpha}^{(n)}(x^\alpha)\, e^{\mathrm{i} n \theta}\,.
    \label{eq:fourier_A}
\end{equation}
We know from \cref{eq:FLTanomaly} that the 't Hooft anomaly takes the form of a total $\theta$-derivative, allowing us to pick out the anomaly-free continuity equation of the zero mode by integrating %
\begin{equation}
    \partial_\alpha j_{(0)}^\alpha(x^\alpha)=\frac{1}{2\pi}\int_{S^1} d\theta\,\partial_\alpha j^\alpha(x^\alpha,\theta)=0\,,
    \label{eq:conserved_integrated_current}
\end{equation}
where the zero is due to integrating the $\theta$-derivative over a closed manifold. Thus the zero-mode $j_{(0)}^\alpha$, describing the U(1) current, is anomaly-free, as it should be since the total current is conserved. Being anomaly-free, we are free to gauge the U(1)-subgroup, meaning that $j_{(0)}^\alpha$ is now the current of a gauge symmetry. The dynamics of the $A_{\alpha}^{(0)}$ that minimally couples to $j_{(0)}^\alpha$, are given by effective Maxwell's equations \cite{gran_holographic_2018};
\begin{equation}
    \partial_\beta {F}_{(0)}^{\beta\alpha}(x^\alpha) - e^2 j^\alpha_{(0)}(x^\alpha) \;=\; 0\,,
    \label{eq:dynamicBC}
\end{equation}
where ${F}_{(0)}^{\beta\alpha}(x^\alpha)$ is the field strength of $A_{\alpha}^{(0)}$ and $e$ is a dimensionless coupling constant which mainly scales the plasma frequency. Since we are only concerned with the existence and qualitative features of the plasmon mode, we will fix this coupling to be $e=1$. 

In the holographic dual bulk theory we only have the anomaly-free LU(1) gauge field as given in \cref{eq:bulk action}. 
In analogy with \cref{eq:fourier_A}, we likewise expand the bulk $A_\mu$ to obtain an infinite tower of gauge fields $A_{\mu}^{(n)}$, $n\in\mathbb{Z}$.

According to the holographic dictionary, the boundary value of each Fourier mode of the bulk gauge field acts as the sources in \cref{eq:fourier_A} for the corresponding boundary current. The boundary conditions for the bulk gauge field are therefore fixed as follows: to obtain a boundary theory with a gauged U(1)-charge, the bulk gauge field zero mode $A_{\mu}^{(0)}$ should obey \cref{eq:dynamicBC} as its conformal boundary condition, while the remaining $A_{n \neq 0, \mu}$ enjoy homogeneous Dirichlet boundary conditions. The standard case of non-dynamical electromagnetism can be obtained by setting Dirichlet boundary conditions for $A_{\mu}^{(0)}$ as well.

In order to implement independent boundary conditions for the zero mode and the higher Fourier modes of the gauge field, we must also Fourier expand the bulk equations of motion and the horizon boundary conditions along the internal $S^1$ direction. Focusing on the response of the total, conserved current, means sourcing only the perturbations in the zero mode sector.
This perturbation then propagates through nontrivial Chern-Simons mediated dynamics to the higher Fourier modes, with the mode mixing controlled by the $\theta$-dependence of the background Fermi surface. In \cref{app:numerics} we verify that the amplitudes of higher modes decay exponentially, allowing us to consistently truncate the system to the first $N$ modes. This results in a finite system of $4(2N+1)$ coupled second-order linear ordinary differential equations in the radial coordinate. Further details on the numerical implementation can be found in \cref{app:numerics}.
\begin{figure}[t]
    \centering
    \includegraphics[width=1\linewidth]{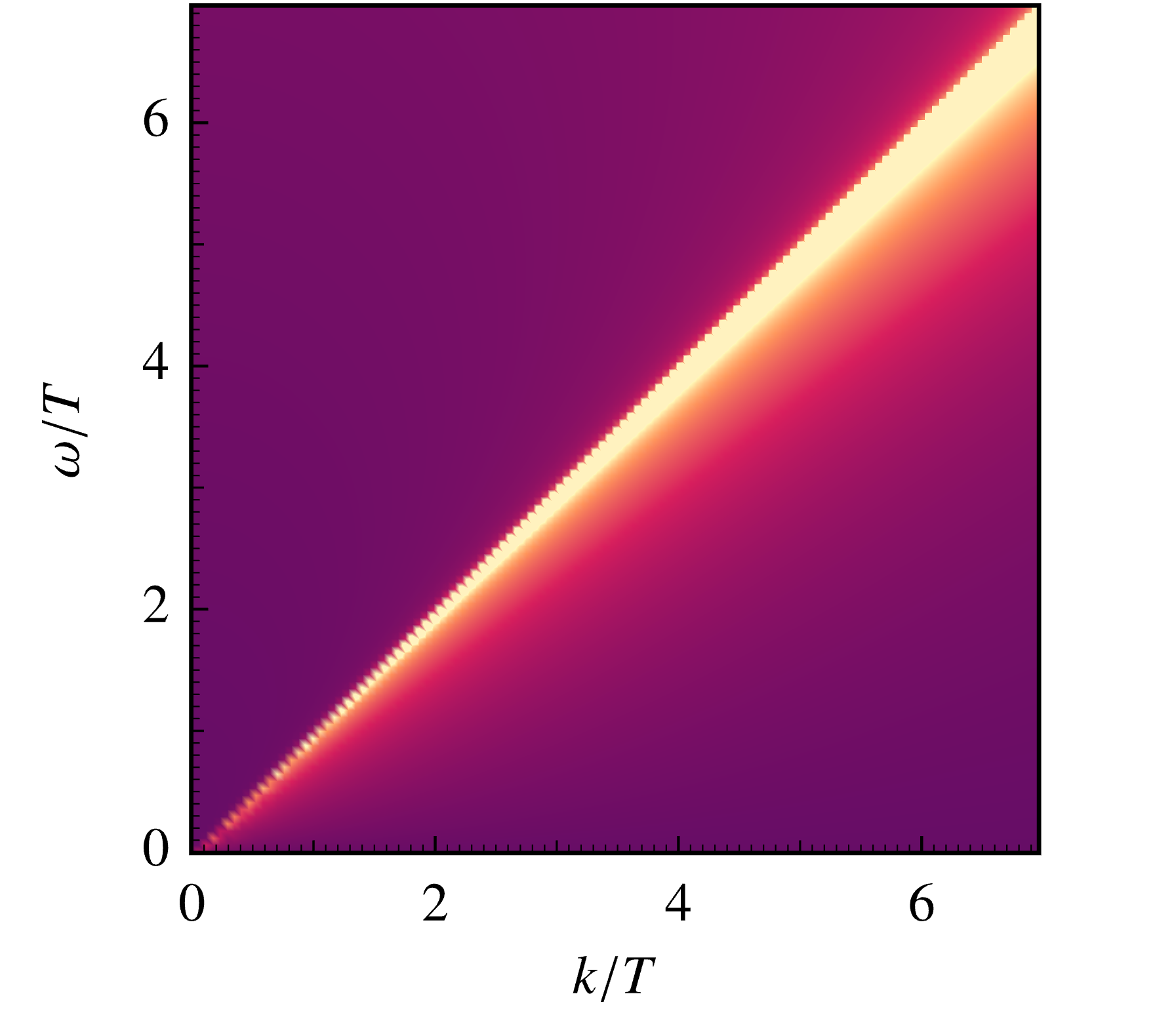}
    \caption{Density-density correlation function for the holographic $\mathrm{LU(1)}$ metal with a circular Fermi surface, for a non-dynamical boundary charge,
    i.e., in the absence of long-range Coulomb interactions. Brighter colors indicate larger values. The spectral weight forms a linearly dispersing feature along $\omega = v_{\mathrm F} k$ (with $v_{\mathrm F}=1$), which broadens into a continuum with support for $\omega \le v_{\mathrm F} k$.}
    \label{fig:Dirrhorho}
\end{figure}
\section{Results}\label{sec:Results}
We begin with the analysis in the absence of long-range Coulomb interactions, i.e., where the gauge field $A_{\mu}^{(0)}$ is treated only as an external source for the current. This corresponds to a non-dynamical total boundary charge, and amounts to imposing Dirichlet boundary conditions also in the zero mode sector.

For concreteness, we will consider the holographic $\mathrm{LU(1)}$ metal with a circular Fermi surface \footnote{Our machinery allows us to consider a Fermi surface of any (closed) shape parameterized by $\theta$. Here we consider the simplest case of a circular Fermi surface, thereby speeding up the numerics.},
\begin{equation}
    (k_x(\theta),k_y(\theta)) = k_{\mathrm F}(\cos\theta,\sin\theta)\,,
\end{equation}
where, as a consequence of considering the IR theory, the Fermi momentum $k_{\mathrm F}$ is parametrically larger than all other scales. Note that the Fermi velocity of the boundary theory is given by the speed of light in the bulk dual, which we set to unity, i.e., $v_{\mathrm F}=c=1$. 

\begin{figure}[b]
    \centering
    \includegraphics[width=1\linewidth]{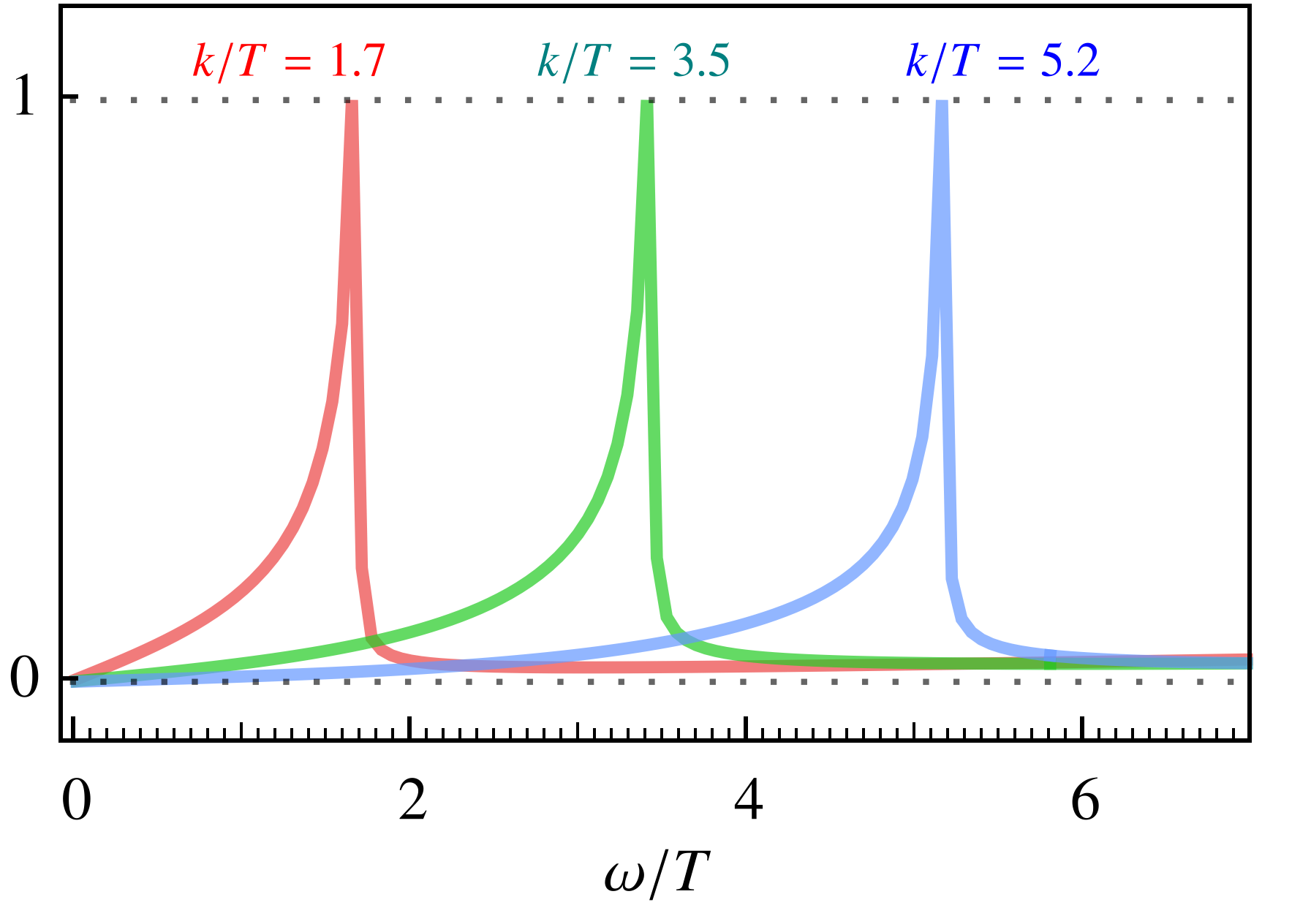}
    \caption{Normalized frequency slices of the density-density correlation function $-\mathrm{Im}\langle \rho\rho\rangle(\omega)$ at fixed momenta $k$. The peaks follow the linear dispersion $\omega = v_{\mathrm F} k$, corresponding to a zero-sound-like mode. The asymmetric broadening reflects Landau-like damping into a particle-hole pseudo-continuum with support $\omega \le v_{\mathrm F} k$ and the quantum critical damping associated with the horizon alone for $\omega \ge v_{\mathrm F} k$.}
    \label{fig:Dirrhorhoslice}
\end{figure}

\Cref{fig:Dirrhorho} shows the density-density correlation function as obtained in this setting. The spectral function exhibits a linearly dispersing peak along $\omega = v_{\mathrm F}k$, which we interpret as zero-sound-like collective mode with a velocity ${v_\mathrm{s}=v_\mathrm{F}}$. It is, however, distinct from the usual neutral conformal zero sound mode $\omega = (c/\sqrt{d}) \, k$ of AdS-Schwarzschild, which resides in the metric sector. It is therefore most plausibly an intrinsic collective excitation of the LU(1) charge sector.
This mode is significantly broadened, with spectral weight extending into a continuum for ${\omega \leq v_{\mathrm F}k}$.

To interpret the continuum we make the following identification: Our choice of units has fixed $v_\text{F} = 1$, and by alluding to a quasiparticle-like picture, $v_\text{F}=k_\text{F}/m^*$. The effective mass $m^*$ of the conserved charges are thus of magnitude $k_\mathrm F$, i.e., also parametrically larger than other scales.
For comparison, the Lindhard particle-hole continuum for a quasiparticle
system with parabolic dispersion has support $\omega \leq v_\text{F}k + k^2/2m^*$~\cite{giuliani_quantum_2005}. In the regime $k \ll k_{\mathrm F}$ relevant to the effective theory, the quadratic term is subleading and the continuum is bounded by $\omega \leq v_{\mathrm F} k$, in agreement with the behavior observed in \cref{fig:Dirrhorho}. This is noteworthy, since a defining feature of holographic models is the lack of quasiparticles. Nevertheless, a broadened continuum resembling the Lindhard particle-hole continuum appears. This suggests that certain qualitative features of such a continuum may persist beyond quasiparticle descriptions and instead reflect more general Fermi-surface kinematics. We note also that this represents a concrete realization of the predicted collective spectrum of EFLs given in \cite{elseCollisionless2023}.

In addition, the holographic model also contains a quantum critical continuum associated with the black hole horizon \cite{romero-bermudez_anomalous_2019}, which normally is the main source of dissipation. The contribution from this sector can be seen more clearly in \cref{fig:Dirrhorhoslice}, where we show slices of constant momentum of the density-density response in \cref{fig:Dirrhorho}. 
The quantum critical continuum provides the small but finite damping in the region $\omega \ge v_{\mathrm F} k$ (right side of each peak) where the Lindhard-like continuum has no support. 

From this explanation, we can attribute the `quasi-Lindhard' continuum 
as coming from the Fermi surface: Since the total current couples to the higher Fourier modes of the currents, it provides additional channels for dissipation of collective excitations in addition to the black hole horizon. Thus, the collective modes of the total charge dissipate into higher modes of the current, which is what gives rise to the observed Landau-like damping. We verify that this is the case by studying the mixed density-density correlation functions $\langle j^t_{(n)} j^t_{(0)} \rangle$ describing the overlap between different modes, which decays with increasing mode number $n$. An explicit computation of this can be found in \cref{app:numerics}.

\subsection{Dynamical electromagnetism}
We now turn to the case of dynamic boundary conditions, corresponding to gauging the anomaly-free $\mathrm{U(1)}$ subgroup and thereby introducing long-range Coulomb interactions.

\begin{figure}[t]
    \centering
    \includegraphics[width=1\linewidth]{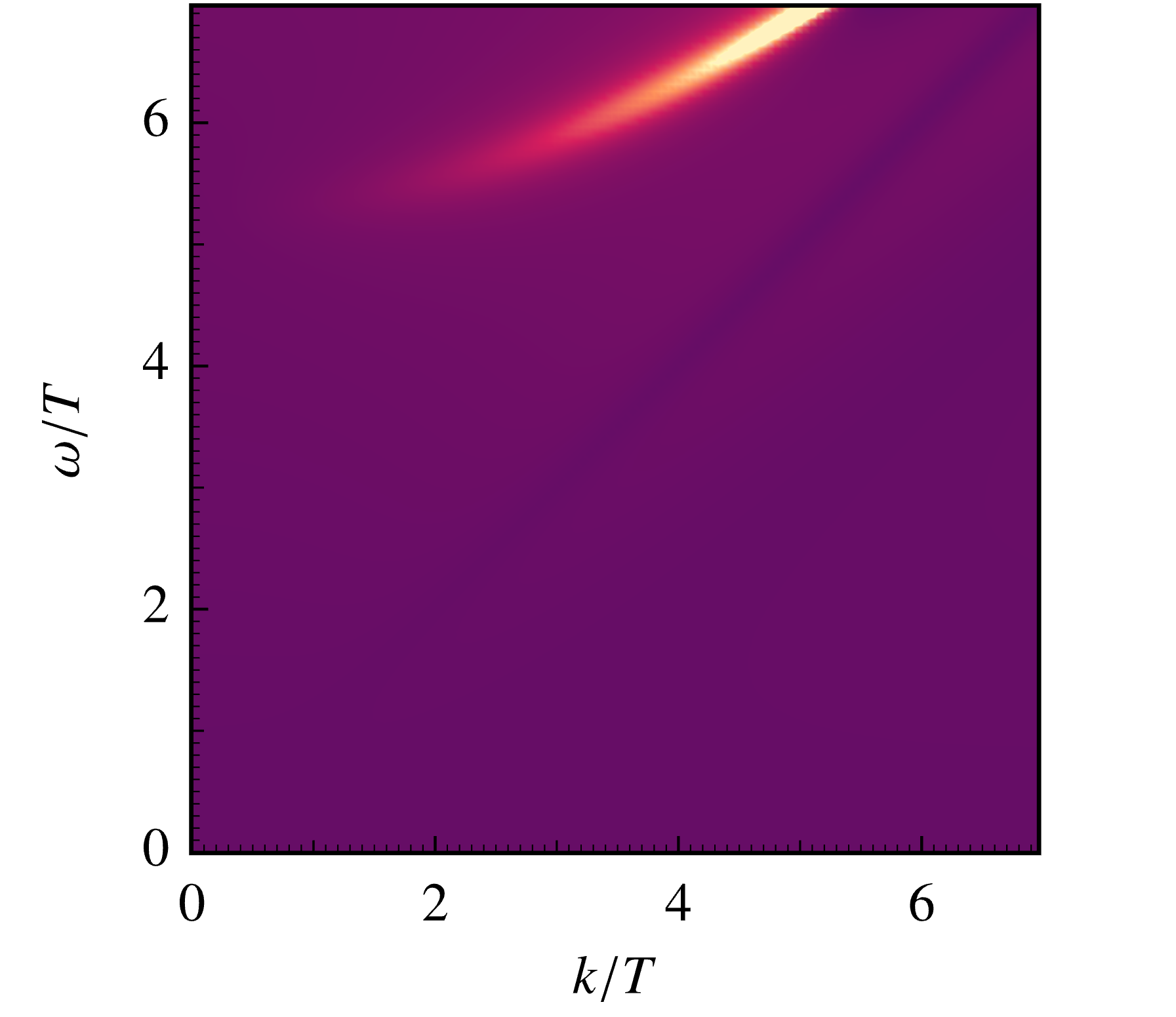}
    \caption{Density-density correlation function for the holographic $\mathrm{LU(1)}$ metal with a circular Fermi surface, evaluated with dynamic boundary conditions corresponding to a gauged $\mathrm{U(1)}$ subgroup. Brighter colors indicate larger values. Long-range Coulomb interactions reorganize the spectral weight into a damped collective mode that is gapped at $k=0$.}
    \label{fig:Dynrhorho}
\end{figure}

The resulting spectral function is shown in \cref{fig:Dynrhorho}.
In contrast to the Dirichlet case, the spectral weight is reorganized into a damped collective excitation. The resulting mode is gapped at zero momentum and exhibits the characteristic dispersion of a plasmon.
This result is quite notable from the holographic point of view:
typically, the plasma frequency $\omega_p$ describing the $k=0$ gap is set by the Drude weight, which in turn is directly proportional to the chemical potential in the bulk \cite{gran_holographic_2018,mauri_screening_2019,romero-bermudez_anomalous_2019}. An AdS-Schwarzschild geometry normally describes zero density matter, which gives $\omega_p=0$. 
In the current setup, charge is instead incorporated through the Fermi surface encoding of the background geometry \cref{eq:metric}.

To further assess the validity of this implementation, we investigate the scaling of the plasma frequency with respect to the volume enclosed by the Fermi surface. From general hydrodynamic results the plasma frequency should scale as $\omega_p^2 \propto\rho/m^*$ in a 2+1-dimensional boundary theory. From our discussion of the Lindhard continuum, recall the relation ${m^* = k_\text{F}}$ due to the identification $v_\text{F} = c$. We conclude that we should have the scaling ${\sqrt{k_\text{F}} \propto \omega_p \propto \sqrt{\rho/m^*} = \sqrt{\rho/k_\text{F}}}$, or equivalently, $\rho\propto k_\text{F}^2\propto\mathcal{V}_\text{F}$, where $\mathcal{V}_\text{F}$ is the volume enclosed by the Fermi surface.

\begin{figure}[b]
    \centering
    \includegraphics[width=1\linewidth]{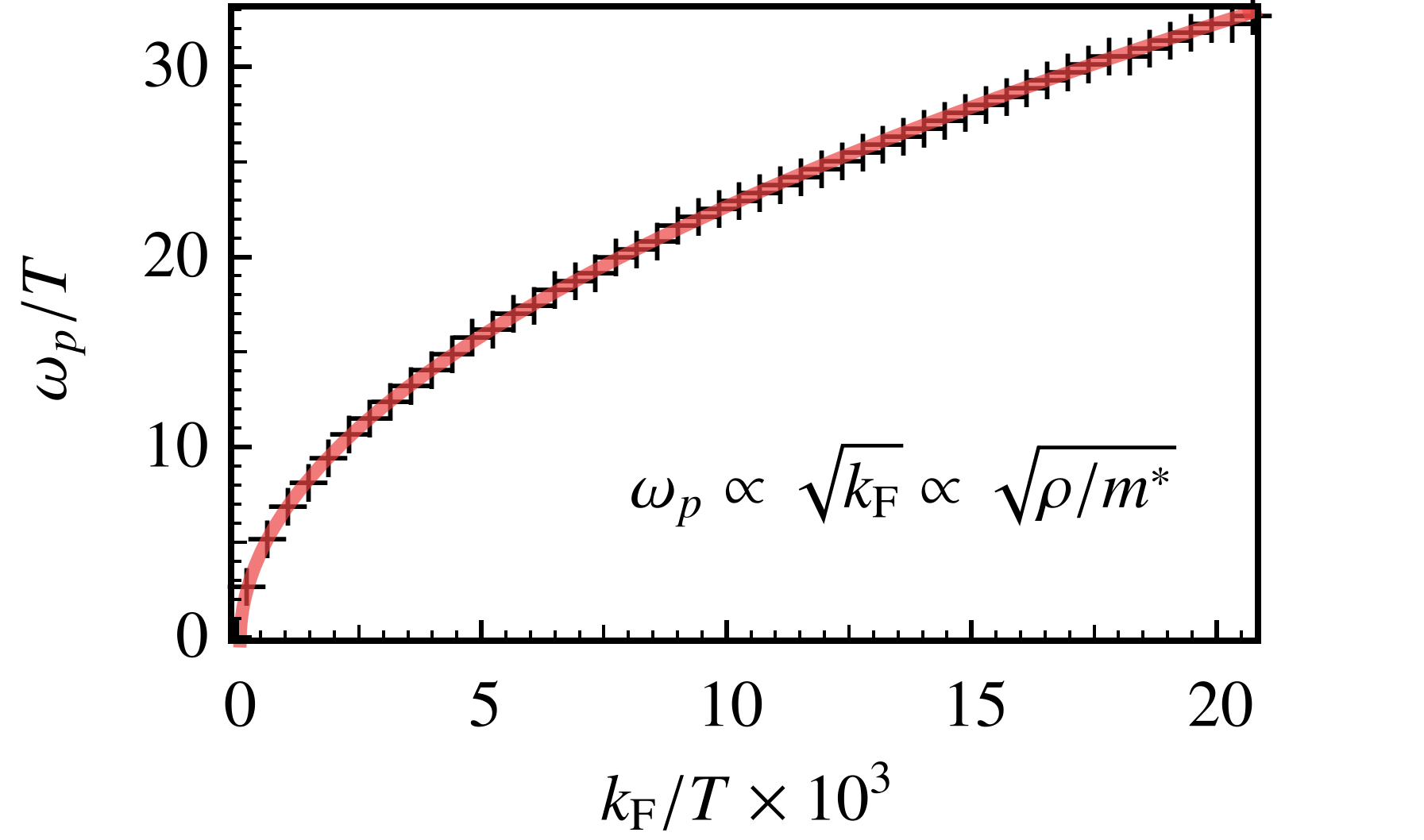}
    \caption{Plasma frequency $\omega_p$ as a function of the Fermi momentum $k_{\mathrm F}$ for the holographic $LU(1)$ metal with dynamic boundary conditions. The numerical data (black markers) are well fit by $\omega_p \propto \sqrt{k_{\mathrm F}}$ (red curve). Given the hydrodynamic expectation $\omega_p^2 \propto \rho/m^*$ together with $m^* \sim k_{\mathrm F}$. This scaling agrees with the plasmon gap predicted from Luttinger’s theorem, $\rho \propto k_{\mathrm F}^2$.}
    \label{fig:plasmaFreqRelation}
\end{figure}

This is indeed the case, as \cref{fig:plasmaFreqRelation} shows. This is the expected scaling from Luttinger's theorem from \cref{eq:generalised-LT} for this holographic model. The fact that this indeed gives rise to a gapped mode serves as a non-trivial check that the EFL framework provides a description of a Fermi surface containing the charge of the electrons of the boundary theory.

Notably, the plasmon mode in \cref{fig:Dynrhorho} lies outside the region ${\omega \leq v_{\mathrm F}k}$, and is therefore protected from the  aforementioned Landau-like damping. Regardless, a characteristic of holographic metals is a strong damping due to dissipation from the black hole horizon \cite{gran_holographic_2018,romero-bermudez_anomalous_2019}.

The neutral black hole contribution persists even in the absence of a Fermi surface. It can be isolated by evaluating the response in the limit $k_\mathrm F \to 0$ where the LU(1) structure collapses to a theory without a Fermi surface and the spectral function is solely the dissipative dynamics of a neutral black hole horizon. The $k_\mathrm F \to 0$ limit corresponds to a well-defined holographic solution in its own right, and thus does not rely on the assumption of $k_\mathrm F$ being parametrically large.

We therefore use the $k_\mathrm F \to 0$ solution as a proxy for the spectrum of the quantum critical background and subtract it from the finite-$k_\mathrm{F}$ result,
\begin{equation}
    \mathrm{Im}\langle\rho\rho\rangle(\omega,k;k_\mathrm F) - \mathrm{Im}\langle\rho\rho\rangle(\omega,k;k_\mathrm F=0)~.
\end{equation}
This subtraction does not correspond to a separation of independent sectors, but provides a diagnostic for isolating the behavior of the collective charge dynamics of the Fermi surface. The spectral weight of the black hole dual is removed, while its dissipative effects on the Fermi surface dynamics remains.
\begin{figure}[t]
    \centering
    \includegraphics[width=1\linewidth]{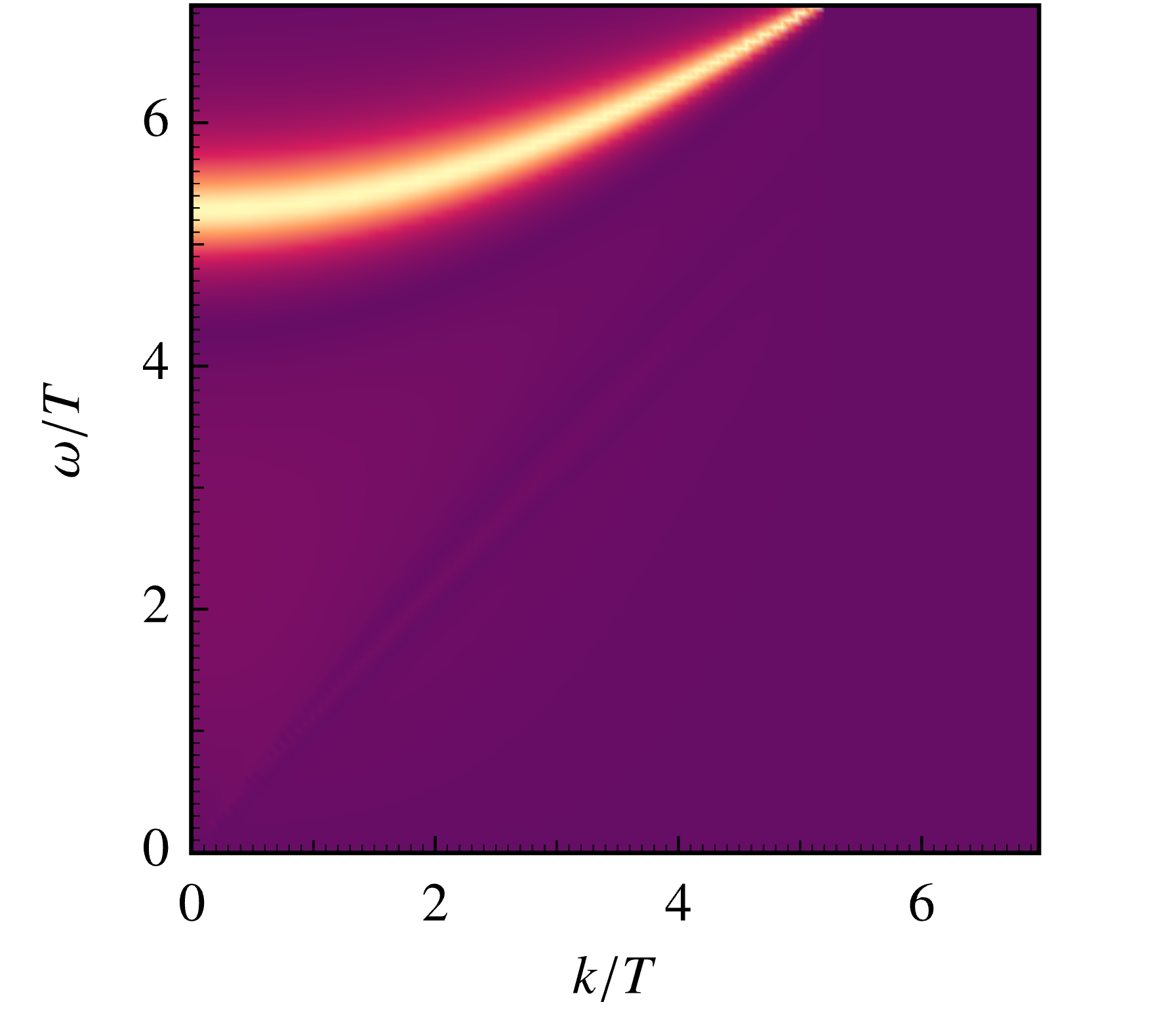}
    \caption{Density-density correlation function with the quantum critical background subtracted and normalized for each $k$. Brighter colors indicate larger values. The plasmon mode is clearly visible and remains gapped at $k=0$ where the linewidth indicates dissipation in this regime.}
    \label{fig:Dynrhorhonorm}
\end{figure}
The response normalized for each $k$ with the spectral weight of the background black hole contribution subtracted is shown in \cref{fig:Dynrhorhonorm}. The residual broadening, most pronounced near $k=0$, can be attributed to quantum critical dissipation from the black hole horizon commonly seen in holographic metals \cite{gran_holographic_2018,romero-bermudez_anomalous_2019}.
\begin{figure}[b]
    \centering
    \includegraphics[width=1\linewidth]{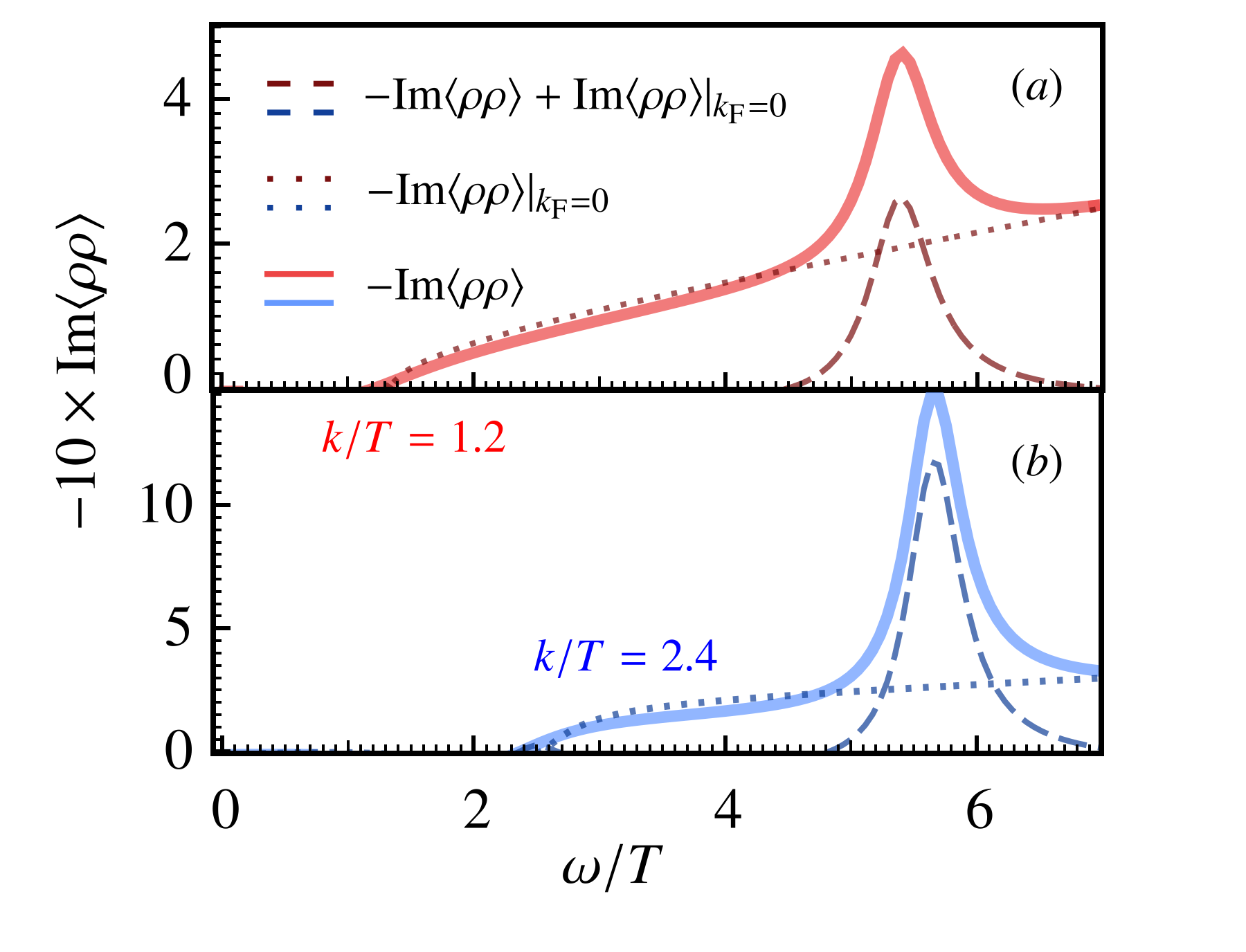}
    \caption{Frequency slices of the density-density correlation function at fixed momenta $k$. The solid curves show the full response, while the dashed curves show the response after subtracting the black hole background. The plasmon appears as a Lorentzian peak, in contrast to the asymmetric line shape of the zero-sound-like mode. The solid curves clearly shows how the plasmon is embedded in the black hole dissipative background. The plasmon can be isolated by removing the background contribution as seen by the dashed lines.}
    \label{fig:Dynrhorhoslice}
\end{figure}
Further insight is obtained by considering frequency slices at fixed momentum, shown in \cref{fig:Dynrhorhoslice}. The full response (solid curves) exhibits a Lorentzian peak embedded in a continuum, reflecting the embedding of the plasmon in the underlying quantum critical background. After subtraction, the resulting peaks (dashed curves) become of Lorentzian lineshape, consistent with a damped collective mode. 

This offers a heuristic explanation for the stronger damping of the plasmon at low momenta. The spectral weight carried by the plasmon scales as $\propto k^2$, whereas the dissipative black hole background remains of comparable magnitude across momenta. At small $k$, the plasmon therefore has relatively little spectral weight compared to this background, leading to strong mixing with the dissipative continuum and consequently significant broadening. As $k$ increases, the plasmon weight grows while the background is of similar magnitude resulting in a sharper plasmon, with reduced relative damping.

We note that this type of response, where a plasmon at low $k$ remains broadened and embedded in a continuum, mirrors experimental observations in strange metals \cite{mitrano_anomalous_2018,chen2024Consistency}.

\section{Conclusion and Outlook}\label{sec:ConclusionAndOutlook}
In this paper, we extend the holographic model of ersatz Fermi liquids \cite{else_holographic_2024} to include dynamical long-range Coulomb interactions, and perform a linear response analysis at finite temperature. Despite the holographically unconventional origin of charge, and the absence of a dynamical coupling to the metric, the explicit inclusion of a Fermi surface through the 't Hooft anomaly gives rise to the collective charge dynamics of a gapped, damped plasmon mode.

The finite damping persists even at zero wavevector, characteristic of previous holographic plasmons \cite{gran_holographic_2018} and argued to be due to a quantum critical continuum in \cite{romero-bermudez_anomalous_2019}. Our results also exhibit qualitative similarities to experimental observations in strange metals, where plasmons are often broad and embedded in a continuum, even at the smallest momenta. \cite{mitrano_anomalous_2018,chen_consistency_2024,vries_reexamining_2026}.

A central result is that the plasma frequency exhibits the expected scaling with the Fermi momentum, consistent with Luttinger’s theorem as encoded kinematically in the LU(1) anomaly structure. This agreement provides a non-trivial consistency check that the ersatz Fermi liquid framework correctly captures the charge density and its collective dynamics, even in the absence of a conventional quasiparticle description. More broadly, it suggests that key aspects of collective charge behavior are already fixed at the level of infrared kinematics, prior to any detailed dynamical modeling.

In the absence of long-range Coulomb interactions, the system instead exhibits a broadened, zero-sound-like mode embedded in a continuum reminiscent of a Lindhard particle-hole spectrum despite the lack of quasiparticles. The emergence of such a structure suggests that aspects of particle-hole kinematics may persist even in the absence of a quasiparticle description, and are captured in the dynamical responses of the EFL framework.

There are several possible directions for future work. It would be valuable to explore more general Fermi surface geometries and their impact on collective modes. Incorporating momentum relaxation or disorder could further bridge the gap to realistic materials. The restriction of the model to be at `zero charge density', in the usual holographic sense, means that it describes a $z=1$ quantum critical theory. This stands in contrast to other successful applications of holography to the strange metal \cite{balm_t-linear_2023,smit_momentum-dependent_2024} where a finite chemical potential induces an AdS$_2\times \mathbb{R}^d$ near-horizon geometry, and thus an emergent IR theory exhibiting local quantum criticality ($z=\infty$). The inclusion of a chemical potential, and hence a dynamical coupling between the graviton and the gauge field, may therefore provide insight into local quantum criticality in holographic EFLs. A further benefit of such an inclusion is that it would necessarily lead to a complete RG-flow encoded in the geometry, whereby standard methods can be used to introduce further irrelevant operators that deform the theory away from the IR fixed point.

Taken together, our results suggest that holographic ersatz Fermi liquids provide a promising route toward a genuinely Fermi-surface-based holographic description of strongly correlated metals. The emergence of a Luttinger-consistent plasmon, together with a Lindhard-like continuum in the absence of long-range Coulomb interactions, indicates that key features of metallic charge dynamics can be encoded directly in the infrared anomaly structure, even without quasiparticles. This makes the EFL framework a natural platform for building more realistic holographic models of strange metals and other non-Fermi-liquid phases. At the same time, the mixed-boundary-condition construction used here to gauge only the anomaly-free subgroup is quite general, and should be useful in other holographic settings where only a non-anomalous sector of a larger symmetry can be made dynamical.

\begin{acknowledgments}
E.N.~is funded by the Nano Area of Advance at Chalmers University of Technology. E.I.~and U.G.~are funded by the KAW grant 2024.0129. The authors would like to thank G.~Ferretti for useful discussions.
\end{acknowledgments}

\crefalias{section}{appsec}
\crefalias{subsection}{appsec}
\appendix
\section{Equations of motion}
\label{app:EOMs}
The linearized bulk equations of motion with the gauge fixing term included,  \cref{eq:BulkEOM_lin_gf}, after making a plane wave ansatz can be written as a matrix equation where
\begin{equation}
    \left( \partial_z^2+\mathcal{M}\partial_z + \Gamma \right)\mathbf{a}=0~,
    \label{eq:MatrixEOM}
\end{equation}
with the notation $\mathbf{a} = (a_t, a_x, a_y, a_z)$ and  $\mathcal{M}$ and $\Gamma$ take the form
\begin{equation}
    \mathcal{M} =
    \begin{bmatrix}
        0 & -\dfrac{k'_y(\theta)}{4\pi^{2}} & \dfrac{k'_x(\theta)}{4\pi^{2}} & 0 \\[1.2ex]
        -\dfrac{k'_y(\theta)}{4\pi^{2} f(z)} & \dfrac{f'(z)}{f(z)} & 0 & 0 \\[1.2ex]
        \dfrac{k'_x(\theta)}{4\pi^{2} f(z)} & 0 & \dfrac{f'(z)}{f(z)} & 0 \\[1.2ex]
        0 & 0 & 0 & \dfrac{2 f'(z)}{f(z)}
    \end{bmatrix}\,.
\end{equation}
and
\begin{widetext}
    \begin{equation}
        \Gamma = 
        \begin{bmatrix}
            -\dfrac{\omega^{2}}{f(z)^{2}}-\dfrac{k^{2}}{f(z)} & 0 & 0 & \dfrac{\mathrm{i}\,k'_{y}(\theta)\,k}{4\pi^{2}} +\dfrac{2\,\mathrm{i}\,\omega}{z} -\dfrac{\mathrm{i}\,\omega\,f'(z)}{f(z)} \\[1.4ex]
            0 & -\dfrac{\omega^{2}}{f(z)^{2}}-\dfrac{k^{2}}{f(z)} & 0 & -\dfrac{\mathrm{i}\,k'_{y}(\theta)\,\omega}{4\pi^{2}f(z)} -\dfrac{2\,\mathrm{i}\,k}{z} \\[1.4ex]
            0 & 0 & -\dfrac{\omega^{2}}{f(z)^{2}}-\dfrac{k^{2}}{f(z)} & \dfrac{\mathrm{i}\,k'_{x}(\theta)\,\omega}{4\pi^{2}f(z)} \\[1.4ex]
            -\dfrac{\mathrm{i}\,k'_{y}(\theta)\,k}{4\pi^{2}f(z)^{2}} + \dfrac{2\,\mathrm{i}\,\omega}{f(z)^{2}z} -\dfrac{\mathrm{i}\,\omega\,f'(z)}{f(z)^{3}} & \dfrac{\mathrm{i}\,k'_{y}(\theta)\,\omega}{4\pi^{2}f(z)^{2}} +\dfrac{2\,\mathrm{i}\,k}{f(z)\,z} & -\dfrac{\mathrm{i}\,k'_{x}(\theta)\,\omega}{4\pi^{2}f(z)^{2}} & -\dfrac{\omega^{2}}{f(z)^{2}} -\dfrac{k^{2}}{f(z)} - \dfrac{2}{z^{2}} +\dfrac{f''(z)}{f(z)}
        \end{bmatrix}\,,
    \end{equation}
\end{widetext}

These are regular-singular equations. To get the regular solutions we first make a Frobenius expansion. Thereafter, we want to specify the conformal boundary conditions for the U(1) subgroup of LU(1) corresponding to gauging the total charge. This is done through a Fourier expansion. The equation can then be solved for the Fourier modes of the regular currents.

\subsection{Horizon boundary conditions}
\label{app:horizonBC}
At finite temperature the bulk geometry contains a black hole horizon that, without loss of generality, we have chosen to be located at $z=1$. In the coordinates \cref{eq:metric}, the linearized equations of motion have a regular singular point at the horizon due to the vanishing of the emblackening factor $f(z)$. Consequently, generic solutions contain both ingoing and outgoing modes.

The regular equations are obtained by imposing ingoing boundary conditions at the horizon. To implement this condition, we perform a Frobenius expansion of the linearized fields about the horizon. Writing
\begin{equation}
    a_\mu(z,\theta) = (1-z)^{\lambda} \sum_{n=0}^{\infty} a^{(n)}_\mu(\theta)\,(1-z)^n ,
\end{equation}
and substituting into the equations of motion fixes the exponents to be
\begin{equation}
    \lambda = \pm \frac{\mathrm{i}\omega}{4\pi T}~,
\end{equation}
where the Hawking temperature is given by
\begin{equation}
    T = \frac{3}{4\pi}~.
\end{equation}
The two branches correspond to outgoing $(+)$ and ingoing $(-)$ modes, respectively. Selecting the ingoing solution therefore amounts to imposing
\begin{equation}
    a_\mu(z,\theta) \sim (1-z)^{-\,\frac{\mathrm{i}\omega}{4\pi T}} \qquad \text{as } z \to 1~.
\end{equation}
Factoring out this universal near-horizon behavior, we write
\begin{equation}
    a_\mu(z,\theta) = (1-z)^{-\,\frac{\mathrm{i}\omega}{4\pi T}} \,\mathring{a}_\mu(z,\theta)~,
\end{equation}
where the rescaled fields $\mathring{a}_\mu$ are required to be analytic at $z=1$. Substituting this form into the equations of motion yields a system that is regular at the horizon and admits a Laurent expansion in $(1-z)$.

The horizon boundary conditions are implemented by imposing that singular leading-order terms in the Laurent expansion obtained from the Frobenius analysis are zero. This results in four Robin boundary conditions at $z=1$, which relate radial derivatives $\partial_z \mathring{a}_\mu$ to the horizon values $\mathring{a}_\mu$. With these conditions imposed, the bulk problem becomes a well-posed boundary value problem. The remaining four degrees of freedom are fixed at the conformal boundary. Solving the resulting system then uniquely determines the bulk response.

\section{Loop group expansion}
\label{app:loopgroup}
This appendix clarifies how the emergent $\mathrm{LU(1)}$ symmetry splits into a global $\mathrm{U(1)}$ subgroup corresponding to total charge and an infinite set of additional infinite-dimensional subgroups. This justifies identifying the Fourier zero mode of the bulk fields with the conserved total charge in the boundary theory. 

The group of interest is $\mathrm{LU(1)} \cong C^\infty (S^1, \mathrm{U(1)})$, describing smooth maps from the circle to U(1). Here we use $\mathrm{U(1)}  \cong S^1 \cong \mathbb{R}/\mathbb{Z}$ interchangeably when suitable. The subgroup of constant loops is isomorphic to $\mathrm{U(1)}$, which gives the canonical decomposition
\begin{equation}
    \mathrm{LU(1)}  \cong \mathrm{U(1)} \times\Omega_1\mathrm{U(1)}~.
    \label{eq:LU1split}
\end{equation}
$\Omega_x\mathrm{U(1)}$ here denotes the \textit{based} loop group, meaning loops $\mathbb{R}/\mathbb{Z}\to \mathrm{U(1)}$ such that $0\mapsto x\in \mathrm{U(1)}$, where in \cref{eq:LU1split} we have chosen $x=1$. 

Identify $\mathrm{U(1)} \cong \mathbb{R}/\mathbb{Z}$ and consider the cover $\mathbb{R}$ of $\mathbb{R}/\mathbb{Z}$. There exists a unique lift \cite{janssen2016LoopGroups} from $\mathbb{R}/\mathbb{Z}$ to $\mathbb{R}$ allowing us to construct another group isomorphism as
\begin{equation}
    \begin{aligned}
        \mathrm{LU(1)} \cong& \; \mathrm{U(1)}\times\bigsqcup_{k\in\mathbb{Z}} \Omega^{(k)}_1\mathrm{U(1)} \\
        \cong& \; \mathrm{U(1)} \times \Omega^{(0)}_1\mathrm{U(1)}\times \mathbb{Z}~,
        \label{eq:fullLU1splitapp}
    \end{aligned}
\end{equation}
where the $k\in\mathbb{Z}$, or equivalently the group $\mathbb{Z}$, specifies the winding number of loops around U(1).

Next, we analyze the Lie algebra. It can be identified from the tangent space at the identity of the codomain of the maps as $\mathrm{L}\mathbb{R}\cong C^\infty (S^1, \mathbb{R})$ \cite{pressley1988Loop}. The algebra, much like the group, allows for a split into constant maps and based maps as
\begin{equation}
    \mathrm{L}\mathbb{R} \cong \mathbb{R}\oplus\Omega_0 \mathbb{R}\,.
    \label{eq:Lu1splitapp}
\end{equation}
The $\mathbb{R}$ corresponds to the algebra of the U(1) subgroup seen in \cref{eq:fullLU1splitapp} while $\Omega_0 \mathbb{R}$ corresponds to the zero winding sector $\Omega^{(0)}_1 \mathrm{U(1)}$ with $0\in \mathbb{Z}$. The remaining sectors are disconnected components and cannot be generated from the algebra.

The global symmetry of total charge can be identified with the U(1) found in \cref{eq:fullLU1splitapp}. The anomalous LU(1) currents are Lie algebra dual-valued objects. These are sourced by the bulk LU(1) gauge fields, meaning, the gauge fields are Lie algebra-valued, $\mathrm L\mathbb{R}$. The Lie algebra can be identified with periodic functions,
allowing us to make a Fourier expansion of any $f\in \mathrm L\mathbb{R}$. The expansion coefficients $f_n$ then give a natural correspondence to the subalgebras found in \cref{eq:Lu1splitapp}. The zero mode, $f_0$, spans $\mathbb{R}$ generating the U(1), while the remaining modes are the generators of the zero winding sector of based loops $\Omega^{(0)}_1 \mathrm{U(1)}\times0\hookrightarrow\Omega^{(0)}_1 \mathrm{U(1)}\times\mathbb{Z}$.

\section{Numerical implementation}
\label{app:numerics}

In light of \cref{app:loopgroup}, there is a final modification to the equations of motion before solving them numerically. We make a Fourier expansion 
\begin{equation}
    \mathring{a}_\mu(z,\theta)= \sum_{n \in \mathbb{Z}:|n|\leq N} \tilde a^{(n)}_{\mu}(z) \mathrm e^{in\theta}
\end{equation}
in the Frobenius expanded $\mathring{a}_\mu(z,\theta)$ and solve for the first $N$ modes of $\tilde a_{\mu}^{(n)}(z)$ by solving $4(2N+1)$ coupled linear ordinary differential equations with the boundary conditions of regularity at the horizon and \cref{eq:dynamicBC} (or Dirichlet for all modes) at the conformal boundary. 

To do so, we employ a pseudo-spectral algorithm. Consider the linear equation
\begin{equation}
    \begin{aligned}
        &\hat{\mathcal{L}} \phi_i = 0\\
        & \text{b.c.} \left\{
        \begin{aligned}
            &\phi_i(z = 0) = \ldots\\
            &\phi_i(z = 1) = \ldots
        \end{aligned}
        \right.
    \end{aligned}
    \label{eq: pseudospectral linear equation}
\end{equation}
where $\phi_i$ contains the $4(2N+1)$ fields $\tilde a^{(n)}_{\mu}(z)$. 
We discretize these equations on a Chebyshev-Lobatto grid $z=z_l$, $l=1,\ldots N_\text{CL}$ 
\cite{boyd2001Chebyshev},
where
\begin{equation}
    z_l = \frac{1}{2} \left( 1-\cos\left[ \pi \frac{l-1}{N_\text{CL}-1} \right]\right).
    \label{eq: Chebyshev-Lobatto grid}
\end{equation}
\begin{figure}[t]
    \centering
    \includegraphics[width=1\linewidth]{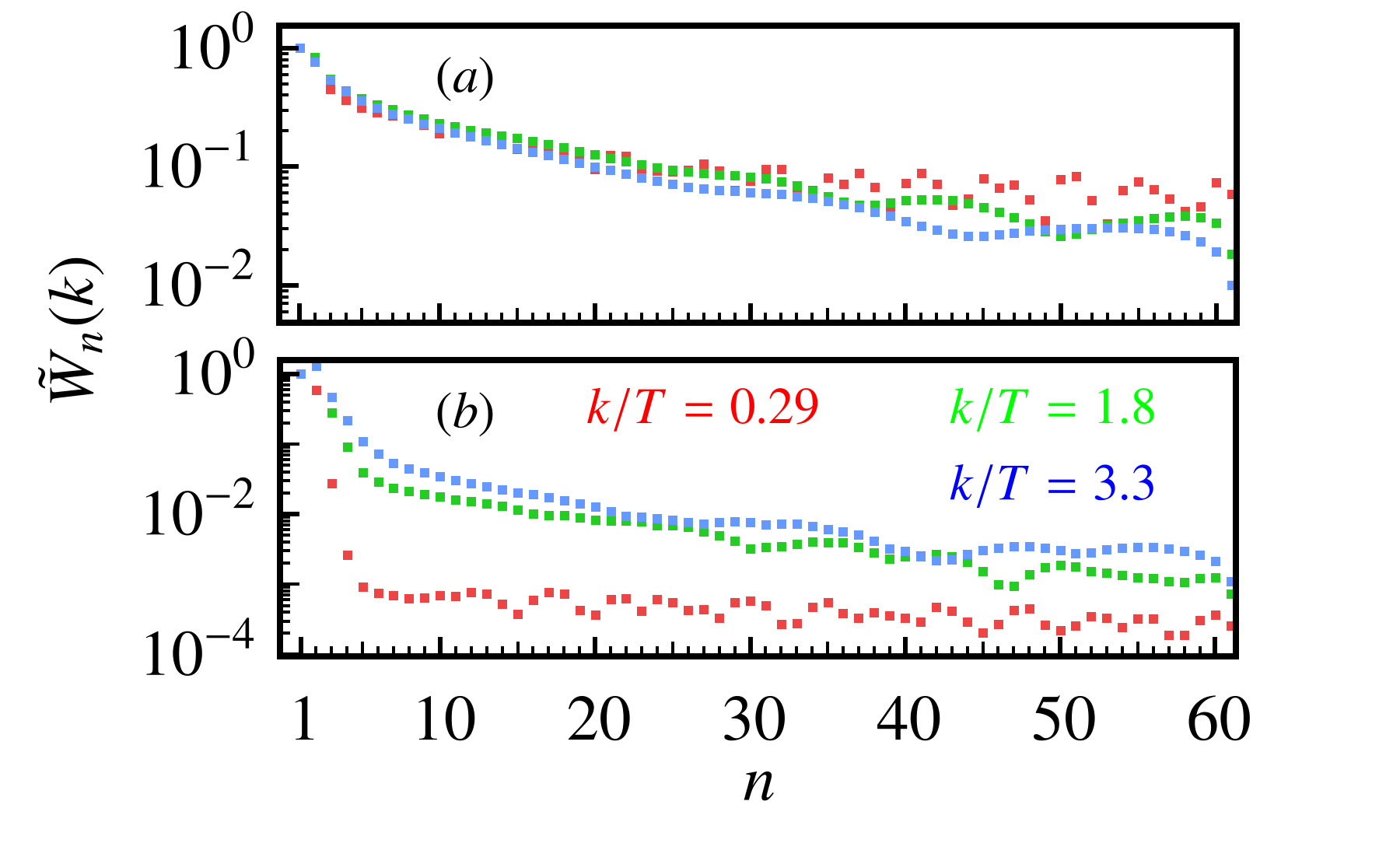}
    \caption{
    Integrated spectral weight of the current-current correlator of \cref{eq:normalisedModeWeight}, as a function of the Fourier mode number $n$, where the integral is done over $I_\omega=[0.01,7.3]$. Results are shown for three values of fixed momenta. Panel $(a)$ shows the response for Dirichlet boundary conditions while panel $(b)$ shows the result for dynamic boundary conditions. Of particular interest is that the spectral weight as a function of mode number for dynamic boundary conditions decay faster, which can be physically understood from the fact that the plasmon does not enter the Lindhard-like continuum, while the zero sound mode does. The spectral weight decays with increasing mode number justifying truncating the loop-group expansion at finite $N=60$ in the numerical implementation. 
    }
    \label{fig:modeweight}
\end{figure}
This discretization turns the Frobenius and Fourier expanded versions of the differential operators in \cref{eq:MatrixEOM} to matrices, whereby the boundary value problem turns into a linear set of equations $A X = b$ which is solved numerically in \texttt{Mathematica}. The boundary currents can be evaluated from the solutions
using the standard holographic correspondence relations in \cref{eq:current}.
Details about further specifics can be found in \cite{ismailov2024Fermi}.

The numerical implementation involves two sources of truncation. First, the loop-group symmetry introduces infinitely many Fourier modes, which we truncate at finite $N$. 
The truncation of the Fourier modes can be quantified by considering the activation of higher Fourier modes under a perturbation of the total charge. We consider the integrated spectral weight
\begin{equation}
    W_n(k) = \int_{I_\omega} \left| \langle j_n^t j_0^t \rangle(\omega,k) \right| \, d\omega,
\end{equation}
where $I_\omega$ is the interval for the integral in $\omega$. We normalize this quantity by its value at $n=0$,
\begin{equation}
    \tilde W_n(k) = \frac{W_n(k)}{W_0(k)}.
    \label{eq:normalisedModeWeight}
\end{equation}
In \cref{fig:modeweight}, $\tilde W_n(k)$ is shown as a function of $n$ for several values of $k$. For Dirichlet boundary conditions, higher Fourier modes carry a larger fraction of the spectral weight compared to the dynamic case. This is consistent with the fact that the zero-sound mode overlaps with the Lindhard-like continuum, whereas the plasmon mode resides in the region $\omega > v_\mathrm{F} k$, where the continuum has no support.

\begin{figure}[b]
    \centering
    \includegraphics[width=1\linewidth]{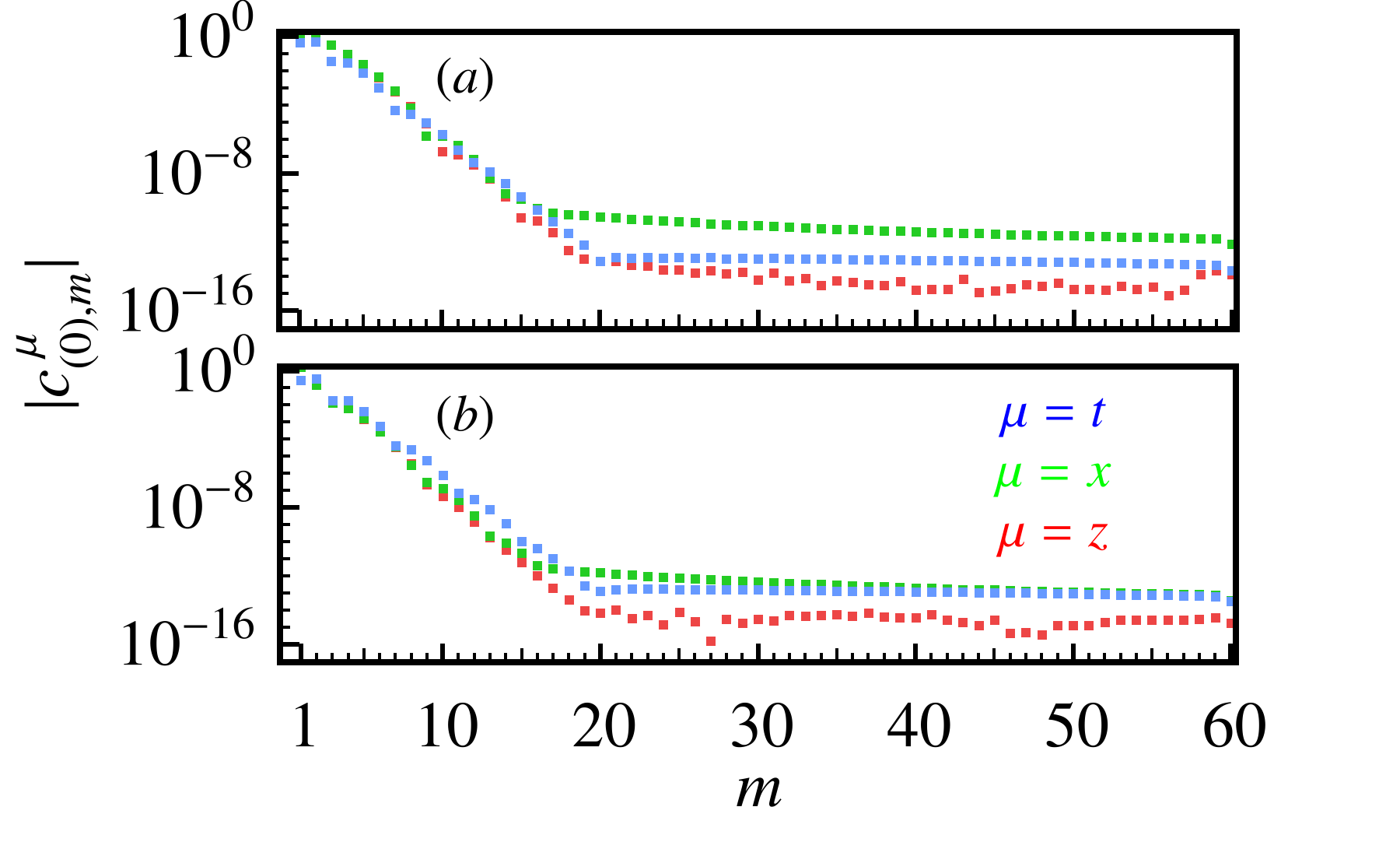}
    \caption{Log-linear plot of the Chebyshev coefficients $|c_{(0),m}^\mu|$ for the Fourier zero mode of the gauge field at fixed ${\omega/T = k/T = 2.5}$. The coefficients for Dirichlet boundary conditions are shown in $(a)$ while the dynamic boundary conditions are shown in $(b)$. Exponential convergence can be seen up until roughly 20 grid points.
    }
    \label{fig:pseudospectralConvergence}
\end{figure}

The pseudo-spectral discretization introduces a second source of truncation controlled by the number of grid points $N_\text{CL}$. As seen in \cref{fig:pseudospectralConvergence}, the Chebyshev coefficients exhibit exponential decay up to $N_\text{CL} \sim 20$, beyond which numerical precision limits further convergence.  The precise number of modes exhibiting exponential decay depends weakly on $\omega$ and $k$, but varies only by a few grid points around this value within our parameter ranges. We therefore use ${N_\text{CL}=35}$ as a conservative truncation for all figures in the main text. Furthermore, we truncate the Fourier expansion at $N=60$, resulting in 484 coupled equations in total.
\nocite{gran2019Holographic, HartnollElectron2011, Hartnoll_2011}
\bibliography{bibliography}

\end{document}

%% file: preamble.tex
\usepackage[dvipsnames]{xcolor}
\usepackage{graphicx}
\usepackage{bm}
\usepackage{comment}

\usepackage[colorlinks=true,
            linkcolor=blue,
            citecolor=blue,
            urlcolor=blue]{hyperref}

\usepackage[nameinlink, english]{cleveref}

\crefname{figure}{Fig.}{Figs.}
\Crefname{figure}{Fig.}{Figs.}

\crefname{table}{Table}{Tables}
\Crefname{table}{Table}{Tables}

\crefname{equation}{Eq.}{Eqs.}
\Crefname{equation}{Eq.}{Eqs.}

\crefname{section}{Sec.}{Secs.}
\Crefname{section}{Sec.}{Secs.}

\usepackage{aliascnt}
\newaliascnt{appsec}{section}
\aliascntresetthe{appsec}
\crefname{appsec}{App.}{Apps.}
\Crefname{appsec}{App.}{Apps.}

%% file: bibliography.bib
@PREAMBLE{
 "\providecommand{\noopsort}[1]{}" 
 # "\providecommand{\singleletter}[1]{#1}%" 
}

@article{else_critical_2021,
  title = {Critical Drag as a Mechanism for Resistivity},
  author = {Else, Dominic V. and Senthil, T.},
  year = {2021},
  month = nov,
  journal = {Physical Review B},
  volume = {104},
  number = {20},
  pages = {205132},
  issn = {2469-9950, 2469-9969},
  doi = {10.1103/PhysRevB.104.205132},
  urldate = {2024-05-29},
  langid = {english}
}

@article{else_non-fermi_2021,
  title = {Non-{{Fermi Liquids}} as {{Ersatz Fermi Liquids}}: {{General Constraints}} on {{Compressible Metals}}},
  shorttitle = {Non-{{Fermi Liquids}} as {{Ersatz Fermi Liquids}}},
  author = {Else, Dominic V. and Thorngren, Ryan and Senthil, T.},
  year = 2021,
  month = apr,
  journal = {Physical Review X},
  volume = {11},
  number = {2},
  pages = {021005},
  issn = {2160-3308},
  doi = {10.1103/PhysRevX.11.021005},
  urldate = {2024-05-13},
  langid = {english}
}

@article{else_strange_2021,
  title = {Strange {{Metals}} as {{Ersatz Fermi Liquids}}},
  author = {Else, Dominic V. and Senthil, T.},
  year = 2021,
  month = aug,
  journal = {Physical Review Letters},
  volume = {127},
  number = {8},
  pages = {086601},
  issn = {0031-9007, 1079-7114},
  doi = {10.1103/PhysRevLett.127.086601},
  urldate = {2024-05-13},
  langid = {english}
}

@misc{else_t_2025,
  title = {'t {{Hooft}} Anomalies in Metals},
  author = {Else, Dominic V.},
  year = 2025,
  month = feb,
  number = {arXiv:2502.19471},
  eprint = {2502.19471},
  publisher = {arXiv},
  doi = {10.48550/arXiv.2502.19471},
  urldate = {2025-04-21},
  archiveprefix = {arXiv}
}

@article{lu_definition_2024,
  title = {Definition and Classification of {{Fermi}} Surface Anomalies},
  author = {Lu, Da-Chuan and Wang, Juven and You, Yi-Zhuang},
  year = 2024,
  month = jan,
  journal = {Physical Review B},
  volume = {109},
  number = {4},
  pages = {045123},
  publisher = {American Physical Society},
  doi = {10.1103/PhysRevB.109.045123},
}

@article{huang2024Effectivea,
  title = {Effective Field Theory for Ersatz {{Fermi}} Liquids},
  author = {Huang, Xiaoyang and Lucas, Andrew and Mehta, Umang and Qi, Marvin},
  year = 2024,
  month = jul,
  journal = {Physical Review B},
  volume = {110},
  number = {3},
  pages = {035102},
  publisher = {American Physical Society},
  doi = {10.1103/PhysRevB.110.035102},
  urldate = {2026-03-08},
}

@article{else_holographic_2024,
  title = {Holographic Models of Non-{{Fermi}} Liquid Metals Revisited: {{An}} Effective Field Theory Approach},
  shorttitle = {Holographic Models of Non-{{Fermi}} Liquid Metals Revisited},
  author = {Else, Dominic V.},
  year = 2024,
  month = jan,
  journal = {Physical Review B},
  volume = {109},
  number = {3},
  pages = {035163},
  issn = {2469-9950, 2469-9969},
  doi = {10.1103/PhysRevB.109.035163},
  urldate = {2024-04-08},
  langid = {english}
}

@article{gran_holographic_2018,
  title = {Holographic Plasmons},
  author = {Gran, U. and Torns{\"o}, M. and Zingg, T.},
  year = 2018,
  month = nov,
  journal = {Journal of High Energy Physics},
  volume = {2018},
  number = {11},
  pages = {176},
  issn = {1029-8479},
  doi = {10.1007/JHEP11(2018)176},
  urldate = {2022-04-09},
  langid = {english}
}

@book{zaanen_holographic_2015,
  title = {Holographic {{Duality}} in {{Condensed Matter Physics}}},
  author = {Zaanen, Jan and Liu, Yan and Sun, Ya-Wen and Schalm, Koenraad},
  year = 2015,
  publisher = {Cambridge University Press},
  address = {Cambridge},
  doi = {10.1017/CBO9781139942492},
  urldate = {2022-04-19},
  isbn = {978-1-139-94249-2},
  langid = {english}
}

@article{romero-bermudez_anomalous_2019,
  title = {Anomalous Attenuation of Plasmons in Strange Metals and Holography},
  author = {{Romero-Berm{\'u}dez}, Aurelio and Krikun, Alexander and Schalm, Koenraad and Zaanen, Jan},
  year = 2019,
  month = jun,
  journal = {Physical Review B},
  volume = {99},
  number = {23},
  pages = {235149},
  issn = {2469-9950, 2469-9969},
  doi = {10.1103/PhysRevB.99.235149},
  urldate = {2022-05-20},
  langid = {english}
}

@article{mitrano_anomalous_2018,
  title = {Anomalous Density Fluctuations in a Strange Metal},
  author = {Mitrano, M. and Husain, A. A. and Vig, S. and Kogar, A. and Rak, M. S. and Rubeck, S. I. and Schmalian, J. and Uchoa, B. and Schneeloch, J. and Zhong, R. and Gu, G. D. and Abbamonte, P.},
  year = 2018,
  month = may,
  journal = {Proceedings of the National Academy of Sciences},
  volume = {115},
  number = {21},
  pages = {5392--5396},
  issn = {0027-8424, 1091-6490},
  doi = {10.1073/pnas.1721495115},
  urldate = {2022-05-12},
  langid = {english}
}

@article{chen_consistency_2024,
  title = {Consistency between Reflection Momentum-Resolved Electron Energy Loss Spectroscopy and Optical Spectroscopy Measurements of the Long-Wavelength Density Response of {{Bi}}{\textsubscript{2}}{{Sr}}{\textsubscript{2}}{{CaCu}}{\textsubscript{2}}{{O}}{\textsubscript{8+x}}},
  author = {Chen, Jin and Guo, Xuefei and Boyd, Christian and Bettler, Simon and Kengle, Caitlin and Chaudhuri, Dipanjan and Hoveyda, Farzaneh and Husain, Ali and Schneeloch, John and Gu, Genda and Phillips, Philip and Uchoa, Bruno and Chiang, Tai-Chang and Abbamonte, Peter},
  year = 2024,
  month = jan,
  journal = {Physical Review B},
  volume = {109},
  number = {4},
  pages = {045108},
  issn = {2469-9950, 2469-9969},
  doi = {10.1103/PhysRevB.109.045108},
  urldate = {2024-01-12},
  langid = {english}
}

@misc{vries_reexamining_2026,
  title = {Reexamining the Strange Metal Charge Response with Transmission Inelastic Electron Scattering},
  author = {de Vries, Niels and Hoglund, Eric and Chaudhuri, Dipanjan and hyun Bae, Sang and Chen, Jin and Guo, Xuefei and Balut, David and Gu, Genda and Huang, Pinshane and Hachtel, Jordan and Abbamonte, Peter},
  year = 2026,
  month = feb,
  number = {arXiv:2602.02348},
  eprint = {2602.02348},
  publisher = {arXiv},
  urldate = {2026-02-03},
  archiveprefix = {arXiv}
}

@article{balm_t-linear_2023,
  title = {T-Linear Resistivity, Optical Conductivity, and {{Planckian}} Transport for a Holographic Local Quantum Critical Metal in a Periodic Potential},
  author = {Balm, F. and Chagnet, N. and Arend, S. and Aretz, J. and Grosvenor, K. and Janse, M. and Moors, O. and Post, J. and Ohanesjan, V. and {Rodriguez-Fernandez}, D. and Schalm, K. and Zaanen, J.},
  year = 2023,
  month = sep,
  journal = {Physical Review B},
  volume = {108},
  number = {12},
  pages = {125145},
  issn = {2469-9950, 2469-9969},
  doi = {10.1103/PhysRevB.108.125145},
  urldate = {2024-03-26},
  langid = {english}
}

@article{smit_momentum-dependent_2024,
  title = {Momentum-Dependent Scaling Exponents of Nodal Self-Energies Measured in Strange Metal Cuprates and Modelled Using Semi-Holography},
  author = {Smit, S. and Mauri, E. and Bawden, L. and Heringa, F. and Gerritsen, F. and Van Heumen, E. and Huang, Y. K. and Kondo, T. and Takeuchi, T. and Hussey, N. E. and Allan, M. and Kim, T. K. and Cacho, C. and Krikun, A. and Schalm, K. and Stoof, H.T.C. and Golden, M. S.},
  year = 2024,
  month = may,
  journal = {Nature Communications},
  volume = {15},
  number = {1},
  pages = {4581},
  issn = {2041-1723},
  doi = {10.1038/s41467-024-48594-6},
  urldate = {2024-06-07},
  langid = {english}
}

@article{Hartnoll_2011,
  title = {Holographically Smeared {{Fermi}} Surface: {{Quantum}} Oscillations and {{Luttinger}} Count in Electron Stars},
  author = {Hartnoll, S. A. and Hofman, D. M. and Tavanfar, A.},
  year = {2011},
  journal= {Europhysics Letters},
  volume = {95},
  number = {3},
  pages = {31002},
  doi = {10.1209/0295-5075/95/31002},
  url = {https://doi.org/10.1209/0295-5075/95/31002}
}

@article{HartnollElectron2011,
  title = {Electron stars for holographic metallic criticality},
  author = {Hartnoll, Sean A. and Tavanfar, Alireza},
  journal = {Phys. Rev. D},
  volume = {83},
  issue = {4},
  pages = {046003},
  numpages = {13},
  year = 2011,
  month = {Feb},
  publisher = {American Physical Society},
  doi = {10.1103/PhysRevD.83.046003},
  url = {https://link.aps.org/doi/10.1103/PhysRevD.83.046003}
}

@article{Polchinski_2012,
  title = {Large-Density Field Theory, Viscosity and `{{2kF}}' Singularities from String Duals},
  author = {Polchinski, Joseph and Silverstein, Eva},
  year = 2012,
  month = aug,
  journal = {Classical and Quantum Gravity},
  volume = {29},
  number = {19},
  pages = {194008},
  publisher = {IOP Publishing},
  doi = {10.1088/0264-9381/29/19/194008},
}

@article{faulknerFriedelHorizon2013,
  title = {Friedel Oscillations and Horizon Charge in {{1D}} Holographic Liquids},
  author = {Faulkner, Thomas and Iqbal, Nabil},
  year = 2013,
  journal = {Journal of High Energy Physics},
  shortjournal = {J. High Energ. Phys.},
  volume = {2013},
  number = {7},
  pages = {60},
  issn = {1029-8479},
  doi = {10.1007/JHEP07(2013)060},
  url = {https://doi.org/10.1007/JHEP07(2013)060},
}

@article{gran2019Holographic,
  title = {Holographic Response of Electron Clouds},
  author = {Gran, U. and Tornsö, M. and Zingg, T.},
  date = {2019-03-05},
  journal = {Journal of High Energy Physics},
  year = 2019,
  shortjournal = {J. High Energ. Phys.},
  volume = {2019},
  number = {3},
  pages = {19},
  issn = {1029-8479},
  doi = {10.1007/JHEP03(2019)019},
  url = {https://doi.org/10.1007/JHEP03(2019)019},
}

@article{iqbalLuttingersTheoremSuperfluid2012,
  title = {Luttinger's Theorem, Superfluid Vortices and Holography},
  author = {Iqbal, Nabil and Liu, Hong},
  year = 2012,
  month = oct,
  journal = {Classical and Quantum Gravity},
  volume = {29},
  number = {19},
  pages = {194004},
  issn = {0264-9381, 1361-6382},
  doi = {10.1088/0264-9381/29/19/194004},
  urldate = {2026-03-05},
  langid = {english}
}

@article{elseCollisionless2023,
  title = {Collisionless Dynamics of General Non-{{Fermi}} Liquids from Hydrodynamics of Emergent Conserved Quantities},
  author = {Else, Dominic V.},
  year = 2023,
  month = jul,
  journal = {Physical Review B},
  volume = {108},
  number = {4},
  eprint = {2301.10775},
  pages = {045107},
  issn = {2469-9950, 2469-9969},
  doi = {10.1103/PhysRevB.108.045107},
}

@book{boyd2001Chebyshev,
  title = {Chebyshev and {{Fourier}} Spectral Methods},
  author = {Boyd, John P.},
  year = 2001,
  publisher = {Courier Corporation},
  urldate = {2026-03-09}
}

@mastersthesis{ismailov2024Fermi,
    author = {Ismailov, Eli},
    title = {Fermi {{Surfaces}} of {{Holographic Metals}}},
    school = {Chalmers University of Technology},
    year = 2024
}

@book{hartnoll_holographic_2018,
  title = {Holographic Quantum Matter},
  author = {Hartnoll, Sean A. and Lucas, Andrew and Sachdev, Subir},
  year = 2018,
  month = mar,
  publisher = {The MIT Press},
  address = {Cambridge},
  urldate = {2023-02-22}
}

@misc{polchinski_effective_1999,
  title = {Effective {{Field Theory}} and the {{Fermi Surface}}},
  author = {Polchinski, Joseph},
  year = 1999,
  month = jun,
  number = {arXiv:hep-th/9210046},
  eprint = {hep-th/9210046},
  publisher = {arXiv},
  urldate = {2023-07-14},
  archiveprefix = {arXiv}
}

@article{shankar_renormalization-group_1994,
  title = {Renormalization-Group Approach to Interacting Fermions},
  author = {Shankar, R.},
  year = 1994,
  month = jan,
  journal = {Reviews of Modern Physics},
  volume = {66},
  number = {1},
  pages = {129--192},
  issn = {0034-6861, 1539-0756},
  doi = {10.1103/RevModPhys.66.129},
  urldate = {2026-03-24},
  copyright = {http://link.aps.org/licenses/aps-default-license},
  langid = {english}
}

@article{mauri_screening_2019,
  title = {Screening of {{Coulomb}} Interactions in {{Holography}}},
  author = {Mauri, Enea and Stoof, Henk},
  year = 2019,
  month = apr,
  journal = {Journal of High Energy Physics},
  volume = {2019},
  number = {4},
  pages = {35},
  issn = {1029-8479},
  doi = {10.1007/JHEP04(2019)035}
}

@article{gynther2011Holographic,
  title = {Holographic Anomalous Conductivities and the Chiral Magnetic Effect},
  author = {Gynther, Antti and Landsteiner, Karl and {Pena-Benitez}, Francisco and Rebhan, Anton},
  year = 2011,
  month = feb,
  journal = {Journal of High Energy Physics},
  volume = {2011},
  number = {2},
  pages = {110},
  issn = {1029-8479},
  doi = {10.1007/JHEP02(2011)110},
}

@article{jimenez-alba2014Anomalous,
  title = {Anomalous Transport in Holographic Chiral Superfluids via {{Kubo}} Formulae},
  author = {{Jimenez-Alba}, Amadeo and Melgar, Luis},
  year = 2014,
  month = oct,
  journal = {Journal of High Energy Physics},
  volume = {2014},
  number = {10},
  pages = {120},
  issn = {1029-8479},
  doi = {10.1007/JHEP10(2014)120},
}

@article{landsteiner2012Holographic,
  title = {Holographic Flow of Anomalous Transport Coefficients},
  author = {Landsteiner, Karl and Melgar, Luis},
  year = 2012,
  month = oct,
  journal = {Journal of High Energy Physics},
  volume = {2012},
  number = {10},
  pages = {131},
  issn = {1029-8479},
  doi = {10.1007/JHEP10(2012)131},
}

@article{bardeen1984Consistent,
  title = {Consistent and Covariant Anomalies in Gauge and Gravitational Theories},
  author = {Bardeen, William A. and Zumino, Bruno},
  year = 1984,
  month = oct,
  journal = {Nuclear Physics B},
  volume = {244},
  number = {2},
  pages = {421--453},
  issn = {0550-3213},
  doi = {10.1016/0550-3213(84)90322-5},
}

@article{luttinger_fermi_1960,
  title = {Fermi {{Surface}} and {{Some Simple Equilibrium Properties}} of a {{System}} of {{Interacting Fermions}}},
  author = {Luttinger, J. M.},
  year = 1960,
  month = aug,
  journal = {Physical Review},
  volume = {119},
  number = {4},
  pages = {1153--1163},
  issn = {0031-899X},
  doi = {10.1103/PhysRev.119.1153},
  urldate = {2026-04-15},
  copyright = {http://link.aps.org/licenses/aps-default-license},
  langid = {english}
}

@misc{janssen2016LoopGroups,
  title = {Loop Groups},
  author = {Janssen, Bas},
  year = {2016},
  note = {Lecture notes from a minicourse at the Summer School on Geometry, Utrecht},
  url = {https://www.bjadres.nl/}
}

@book{pressley1988Loop,
  title = {Loop {{Groups}}},
  author = {Pressley, Andrew and Segal, Graeme and Pressley, Andrew and Segal, Graeme},
  year = 1988,
  month = jun,
  series = {Oxford {{Mathematical Monographs}}},
  publisher = {Oxford University Press},
  address = {Oxford, New York},
}

@book{giuliani_quantum_2005,
  title = {Quantum Theory of the Electron Liquid},
  author = {Giuliani, G. F. and Vignale, G.},
  year = 2005,
  publisher = {Cambridge University Press},
  address = {New York}
}

@article{chen2024Consistency,
  title = {Consistency between Reflection Momentum-Resolved Electron Energy Loss Spectroscopy and Optical Spectroscopy Measurements of the Long-Wavelength Density Response of \$\textbraceleft\textbackslash mathrm\textbraceleft{{Bi}}\textbraceright\textbraceright\_\textbraceleft 2\textbraceright\textbraceleft\textbackslash mathrm\textbraceleft{{Sr}}\textbraceright\textbraceright\_\textbraceleft 2\textbraceright\textbraceleft\textbackslash mathrm\textbraceleft{{CaCu}}\textbraceright\textbraceright\_\textbraceleft 2\textbraceright\textbraceleft\textbackslash mathrm\textbraceleft{{O}}\textbraceright\textbraceright\_\textbraceleft 8+x\textbraceright\$},
  author = {Chen, Jin and Guo, Xuefei and Boyd, Christian and Bettler, Simon and Kengle, Caitlin and Chaudhuri, Dipanjan and Hoveyda, Farzaneh and Husain, Ali and Schneeloch, John and Gu, Genda and Phillips, Philip and Uchoa, Bruno and Chiang, Tai-Chang and Abbamonte, Peter},
  year = 2024,
  month = jan,
  journal = {Physical Review B},
  volume = {109},
  number = {4},
  pages = {045108},
  publisher = {American Physical Society},
  doi = {10.1103/PhysRevB.109.045108},
}
